\documentclass[useAMS,usenatbib]{mn2e}

\usepackage{graphicx}
\usepackage{hyperref}

\providecommand{\adsurl}[1]{\href{#1}{ADS}}

\usepackage{amssymb,amsmath}

\usepackage{times}

\usepackage{subfig}
\newcommand{\gadget}{{\footnotesize GADGET}}
\newcommand{\HII}{\text{HII}}
\def\zini{z_{ini}}

\title{The High-$z$ Universe Confronts Warm Dark Matter: Galaxy Counts, Reionization and the Nature of Dark Matter}
\author[{Christian Schultz, Jose O\~{n}orbe, Kevork N.\ Abazajian, James S.\ Bullock}]
{Christian Schultz$^{1,2}$\thanks{Email: cs06@phys.au.dk}, Jose O\~{n}orbe$^{1}$, Kevork N.\ Abazajian$^{1}$, James S.\ Bullock$^{1}$
\\
\\
$^1$Center for Cosmology, Department of Physics and Astronomy, 4129 Frederick Reines Hall, University of California, Irvine, CA 92697 \\
$^2$Department of Physics and Astronomy, Aarhus University, Ny Munkegade, DK-8000 Aarhus C, Denmark.
}

\begin{document}
\maketitle

\begin{abstract}
We use $N$-body simulations to show that high-redshift galaxy counts
provide an interesting constraint on the nature of dark matter,
specifically Warm Dark Matter (WDM), owing to the lack of early
structure formation these models.  Our simulations include three WDM
models with thermal-production masses of 0.8 keV, 1.3 keV, and 2.6
keV, as well as CDM.  Assuming a relationship between dark halo mass
and galaxy luminosity that is set by the observed luminosity function
at bright magnitudes, we find that 0.8 keV WDM is disfavored by direct
galaxy counts in the Hubble Ultra Deep Field at $>\!\!10\sigma$.
Similarly, 1.3 keV WDM is statistically inconsistent at $2.2\sigma$.
Future observations with JWST (and possibly HST via the Frontier
Fields) could rule out $1.3$ keV WDM at high significance, and may be
sensitive to WDM masses greater than 2.6 keV.  We also examine the
ability of galaxies in these WDM models to reionize the universe, and
find that 0.8 keV and 1.3 keV WDM produce optical depths to the Cosmic
Microwave Background (CMB) that are inconsistent at 68\% C.L. with
current Planck results, even with extremely high ionizing radiation
escape fractions, and 2.6 keV WDM requires an optimistic escape
fraction to yield an optical depth consistent with Planck
data. Although CMB optical depth calculations are model dependent, we
find a strong challenge for stellar processes alone to reionize the
universe in a 0.8 keV and 1.3 keV WDM cosmology.
\end{abstract}


\section{Introduction}
Dark matter dominates the evolution of gravitational perturbations,
leading to the formation of haloes and galaxies.  In the prevalent
paradigm of cold dark matter (CDM), the primordial perturbation
spectrum extends to very small scales; galaxy formation proceeds from
the bottom up, commencing in the smallest dark matter haloes where gas
cooling can occur.  If instead there exists a non-negligible minimal
scale for primordial perturbations as in the case of warm dark matter
(WDM), halo formation is delayed, and early galaxy formation is
suppressed considerably.

Early galaxy formation has been understood to be a challenge for WDM
models for some time \citep*{Barkana:2001gr,Somerville2003}.  Today,
the tension is only heightened by mounting evidence that structure
formation is proceeding in earnest at very early cosmic times.  There
are now direct detections of galaxies at redshifts as high as $\sim$10
\citep{Ellis2013,Oesch:2013pt}, clearly indicating that there are
collapsed structures at this time.  More indirectly, studies of quasar
spectra show that the intergalactic medium was almost fully ionized by
redshift $z \sim 6$ \citep{Fan2006} and the measured electron
scattering optical depth from the cosmic microwave background may
could imply reionization as early as $z \sim 10$ \citep{Ade:2013zuv}.
The maintenance of reionization back to these early times seems to
require contributions from numerous, low-mass galaxies
\citep{Kistler:2009mv,Kuhlen:2012vy,Robertson:2013bq}.  In this paper,
we examine how current and future observations of high-$z$ galaxies,
together with observational probes of reionization, can constrain the
dark matter power spectrum on small scales, and by extension the
particle nature of dark matter.

There has been considerable interest in the WDM paradigm for galaxy
formation, owing to potential problems with the LCDM model on
sub-galactic scales.  Most recently, it has been recognized that the
observed central densities of low-luminosity Milky Way dwarf satellite
are significantly lower than expected in dissipationless CDM
simulations \citep{BoylanKolchin:2011dk}.  This issue can be be
alleviated if the dark matter is warm
\citep{Lovell:2011rd,Lovell:2013ola,Polisensky2013}.  Here, we
specifically study a WDM model (1.3 keV thermal particle mass) that
corresponds to the cutoff scale that alleviates the central-density
problem. It should be noted here that models including ultra-light
axions alongside a cold dark matter component can also alleviate the
sub-galactic problems in the pure cold dark matter models
\citep{Marsh:2013ywa}.

The two most popular classes of WDM particle candidates are
``thermal'' particles and sterile neutrinos.  Thermal WDM is coupled
to the primordial plasma in the early Universe, and is diluted to the
proper (observed) dark matter density by an unspecified process.
Sterile neutrinos, on the other hand, can be produced at the proper
dark matter abundance through scattering processes due to their mixing
with active neutrinos with the Dodelson-Widrow
mechanism~\citep[][sometimes referred to as non-resonant
  production]{Dodelson:1993je}, through resonant production in the
case of a large cosmological lepton asymmetry \citep{Shi:1998km}, or
through coupling with other fields
\citep{Kusenko:2006rh,Shaposhnikov:2006xi}.  An important
characteristic of the different models is the free streaming length
they introduce, with a given particle mass having a different
free-streaming length for the different WDM particles and for the
different sterile neutrino production mechanisms.  In this paper we
primarily state particle masses in terms of thermal WDM particles,
i.e., the ``thermal mass'', but we also provide conversions to the
Dodelson-Widrow sterile neutrino mass in summary statements and
tables.  We also quote the wave numbers where the associated power
spectra fall to half the value of a standard CDM model, which allows
our results to be interpreted generally for any model that results in
truncated small-scale power, as can arise for standard CDM particles
in the case of non-standard inflation
\citep[e.g.][]{Kamionkowski2000,Zentner2002}.

Recent work has constrained the warm dark particle mass by a number of
methods.  Some of the currently tightest constraints come from
observations of the Lyman-$\alpha$ (Ly$\alpha$) forest produced by
neutral gas along the line of sight to distant quasars.  The neutral
gas follows the gravitationally-dominant dark matter clustering in the
mildly non-linear regime probed by the Ly$\alpha$ forest, and
therefore it can be a powerful probe of the dark matter perturbation
spectrum at small scales.  However, the Ly$\alpha$ forest is a
challenging tool, requiring disentangling the effects of pressure
support and thermal broadening of the Ly$\alpha$ forest features from
the effects of dark matter perturbation suppression from WDM.  In
addition, modeling the dependence on the physics of the neutral gas
requires assumptions of the thermal history of the intergalactic
medium and its ionizing background, which are done as parameterized
fitting functions.  Many of the limitations of the Ly$\alpha$ forest
on constraints of the primordial power spectrum are discussed in
\citet{Abazajian:2011dt}.  Setting aside the limitations of the
method, the Ly$\alpha$ forest provides stringent constraints, with
recent quoted limits at $m_{\rm WDM}>3.3\rm\ keV\ $
\citep[2$\sigma$,][]{Viel:2013fqw}.

The lack of early structure formation in WDM has motivated limits from
the rate of high-$z$ gamma-ray bursts \citep{deSouza:2013wsa}.
Similarly, \citet{Pacucci:2013jfa} utilize strongly lensed
ultra-faint, high redshift galaxies to constrain the particle mass by
halo mass function considerations.  At low redshift, WDM models can be
constrained by studying the abundance of small galaxies.  Work by
\citet{Polisensky:2010rw} and \citet{Lovell:2013ola} uses N-body
simulations of Milky Way sized dark matter haloes and constrains the
particle mass by assuming that the number of simulated dark matter
satellites equals or exceeds the number of observed Milky Way
satellites, and report limits on thermal WDM particle masses of
$m_{\rm WDM}>2.3\rm\ keV$ and $>1.5\rm\ keV$, respectively.  It should
be recognized that constraints from satellite counts are sensitive to
halo-to-halo variation in substructure counts as well as assumed
completeness corrections to the observed Milky Way satellite
luminosity function.  More recently, \citet{Horiuchi2013} have tried
to adequately account for the halo-to-halo scatter and focused on
counts around M31 (which are higher than around the MW at fixed
luminosity) and find $m_{\rm WDM}>1.8\rm\ keV$.

Given the potential systematic problems with known WDM constraints, it
is useful to explore alternative probes.  In the rapidly-evolving
field of high-$z$ galaxy surveys, the Lyman break technique has proven
useful for discovering galaxies and estimating the UV luminosity
function out to redshifts $z\sim 9$, although there are candidates in
the literature at redshifts as high as $z\sim 12$
\citep{Oesch:2013pt,McLure2012,Bouwens:2011xu,Schenker:2012vs,Bouwens2007}. New
Fourier techniques seem promising in finding fainter candidates below
the normally required detection threshold $S/N\sim4.5$
\citep{Calvi:2013ssa}. Furthermore, Lyman break galaxies seem to be
fair tracers of the overall halo population
\citep{Conroy:2005aq}. Thus the UV luminosity function interconnects
with the halo mass function of dark matter, a quantity which is
readily constructed from simulations and from which different dark
matter models can be distinguished.

Additional physical mechanisms may ease the tension between
simulations and observations presented by, e.g., the missing dwarf
galaxies or the too-big-to-fail problem
\citep{BoylanKolchin:2011dk}. One example is the work
\citet{Bovill:2010bz}. Here it is noted that the Milky Way halo may be
populated by fossils of early dwarf galaxies that formed before
reionization, and that these fossils today have very low surface
brightness rendering them outside of current observational
bounds. Therefore, there may indeed be a population of low luminosity
dwarf galaxies near the Milky Way. \citet{Katz:2012nd} expands on this
result, and argues that the bulk of the old globular clusters in the
Milky Way formed in these first (now fossil) dwarf
galaxies. Furthermore, the proto-globular clusters were an important
mode of star formation in these galaxies, and could in fact be the
main driver for reionization.

In this work, we study in detail the effects of WDM models on high-$z$
dark matter halo counts using high-resolution cosmological
simulations, and extend these results empirically to infer the
observable effects on galaxies and reionization.  We compare the
high-$z$ luminosity functions of galaxies recently measured from the
Hubble Ultra Deep Field (HUDF) with the inferred luminosity-function
derived using $N$-body simulations in WDM and CDM cosmologies.  Our
predicted luminosity functions are normalized to match observed bright
galaxy counts using abundance matching. All magnitudes quoted below
are in the AB system.  The same models allow us to study cosmological
reionization in WDM models and compare them to CDM.

\section{Theory and Simulations}
\label{sec:Theory}
\subsection{Power Spectrum}
\begin{figure}
           \includegraphics[width=\columnwidth]{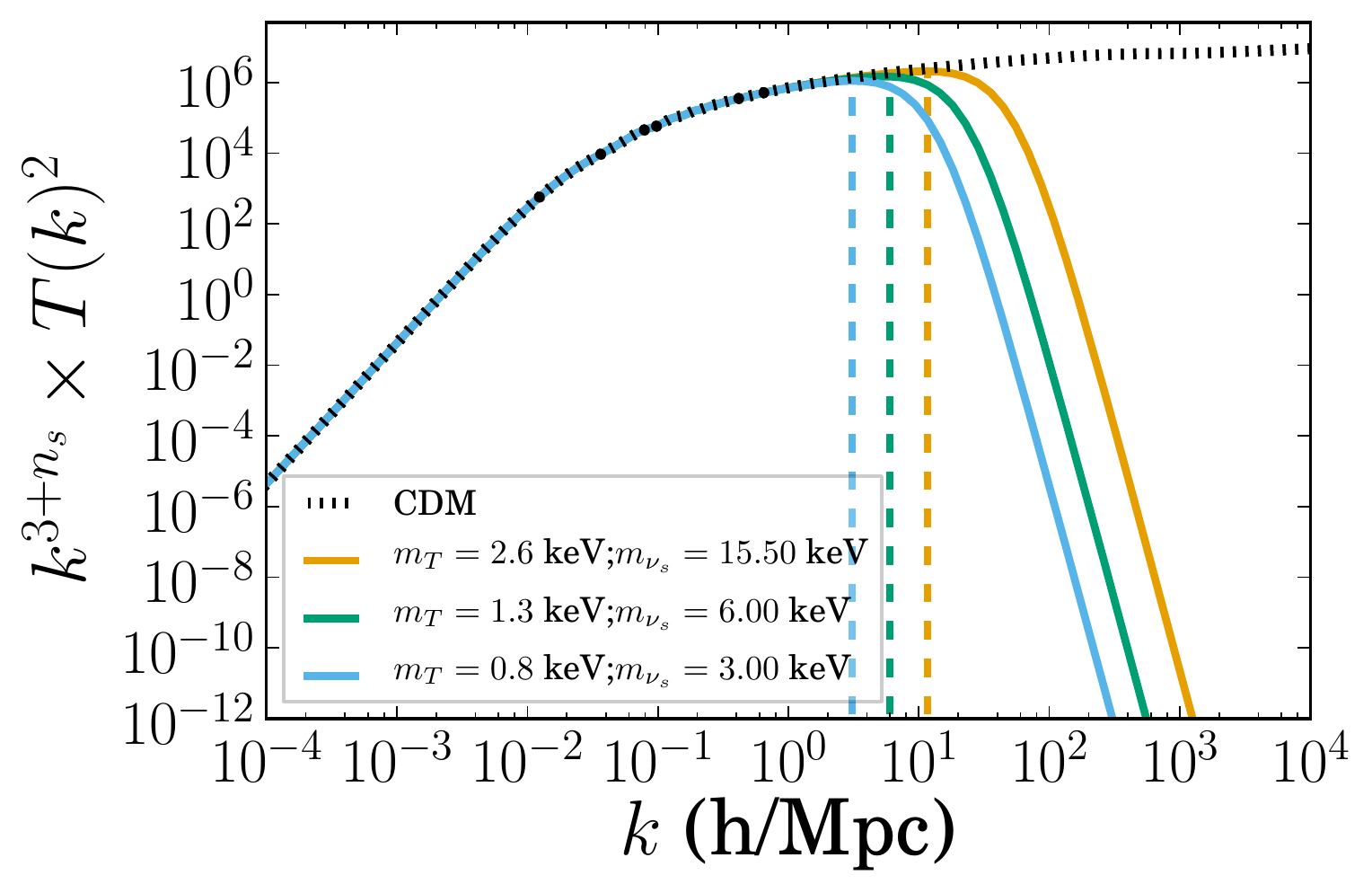}
     \caption{Shown are the matter power spectra for the different WDM
       models considered in this paper. The more massive the WDM
       particle, the more CDM-like the matter transfer function
       becomes due to the lower thermal velocities. For reference,
       each model is labeled by its equivalent thermal and sterile
       neutrino mass.  The dashed lines mark the maximum of the
       transfer function.}
   \label{fig:Transferfuncs}
\end{figure}

WDM has a non-negligible thermal velocity which imprints a free
streaming scale in the matter perturbation distribution arising from
the early Universe. Below this free-streaming scale, structure
formation is suppressed.  This scale is conveniently parameterized by
the Jeans mass
\begin{equation}
M_J(z)=\frac{4}{3\pi}\rho_{\rm dm}(z) \left(\frac{\pi\sigma_v^2(z)}
{4G\bar{\rho}(z)} \right)^{3/2} \, ,
\end{equation}
where $\rho_{\rm dm}$ is the dark matter density, $\bar{\rho}$ the
mean density and $\sigma_v$ the velocity dispersion. The Jeans mass is
constant approximately until matter-radiation equivalence and thus
erases the initial conditions below this mass scale. After
matter-radiation equality the Jeans mass drops rapidly and decays with
the cooling of the dark matter in the Hubble flow as $\sim a^{-3/2}$
\citep{Schneider:2013ria}.  The transfer function relates the
primordial matter power spectrum to the linear power spectrum at a
later redshift. For WDM, the matter power spectrum can be seen as a
suppression of power above a certain wavenumber $k$.  In fact, the
transfer function for WDM relative to the CDM case can be approximated
by the fitting function \citep{Abazajian:2005gj}:
\begin{equation}
T(k)=\left[1+(\alpha k)^{\nu}\right]^{-\mu} \, ,
\label{eq:Transferfunc1}
\end{equation}
where the smoothing scale is set by
\begin{equation}
\alpha=a\left(\frac{m_s}{\mbox{keV}}\right)^b\left(\frac{\Omega_{\rm
    dm}}{0.26}\right)^c\left(\frac{h}{0.7}\right)^d h^{-1}\mbox{ Mpc}
.
\label{eq:Transferfunc}
\end{equation}
Here $m_s$ is the (non-thermally produced) sterile neutrino WDM
particle mass \citep{Abazajian:2005gj}.  The relationship between the
mass of a thermal particle ($m_{\rm WDM}$) and the mass of the sterile
neutrino ($m_{s}$) for which the transfer functions are nearly
identical \citep{Viel:2005qj} is:
\begin{equation}
m_{\rm WDM}=0.335\ {\rm keV} \left(\frac{m_s}{\mbox{keV}}\right)^{3/4}
\left(\frac{\Omega_{m}}{0.266}\right)^{1/4}
\left(\frac{h}{0.71}\right)^{1/2}.
\label{eq:mrel}
\end{equation}
Fig.~\ref{fig:Transferfuncs} shows the matter power spectrum for the
three WDM models we explore in this paper (labeled by thermal mass and
equivalent sterile mass) along with a CDM model for comparison.  For
comparison purposes, the scales where these WDM transfer functions
equal half the CDM transfer function, $k_{1/2}$, are listed in Table
\ref{tab:Transferfuncs}, along with wave number of maximum power
$k_{\rm max}$.

\begin{figure*}
     \begin{centering}
\includegraphics{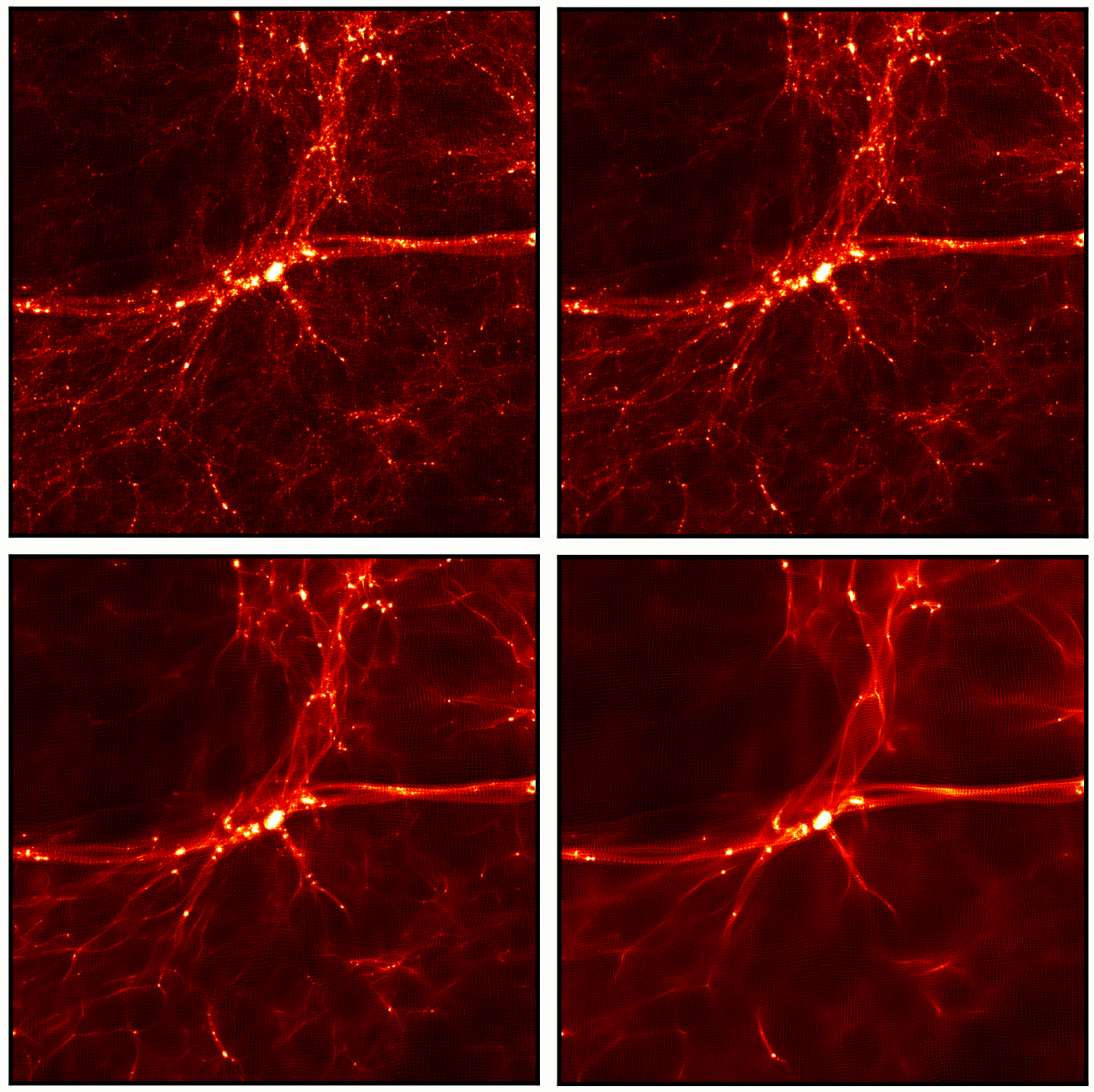}
     \end{centering}
\caption{Simulation images for CDM and WDM at z=6, each initialized
  with the same random seed.  The panels are $10 h^{-1}$Mpc square and
  6 $h^{-1}$Mpc deep; they are centered on the most massive halo in
  the box.  The upper left panel is CDM, with 2.6 keV WDM in the upper
  right.  The bottom panels correspond to WDM: 1.3 keV (left) and 0.8
  keV (right).  The lack of structure for the lightest WDM models is
  striking compared to CDM.}
  \label{fig:snapshots} 
\end{figure*}

\subsection{Numerical Simulations}
\label{sec:Nbody}
Our simulations were performed with the \gadget-2 code, in TreePM mode
\citep{Springel:2005mi}. In order to generate the initial conditions
(ICs), we have used the MUSIC code \citep{Hahn:2011}. The method uses
an adaptive convolution of Gaussian white noise with a real-space
transfer function kernel together with an adaptive multi-grid Poisson
solver to generate displacements and velocities following second-order
Lagrangian perturbation theory. For more specific details on the MUSIC
code, we refer the reader to \citep{Hahn:2011}.  The \texttt{CAMB}
\citep{Lewis:2000,Howlett:2012} package was used to generate the CDM
transfer functions used to generate the ICs for this cosmology. The
WDM transfer functions were obtained from CDM using equations
\eqref{eq:Transferfunc1}-\eqref{eq:Transferfunc}. Only the initial
conditions were modified, and the thermal velocities of the WDM
particles are not included in the simulations, since they have not
been found to be significant in affecting WDM structure formation
\citep{Bode:2000gq,VillaescusaNavarro:2010qy}.

The cosmological parameters used were $h=0.71$, $\Omega_m=0.266$,
$\Omega_\Lambda=0.734$, $n_s=0.963$ and $\sigma_8=0.801$. All
simulations are $1024^3$ particles in (50 Mpc/$h$)$^3$ boxes started
at $\zini=125$.  The implied particle mass is $m_p = 8.6 \times 10^6
h^{-1}M_\odot$.  We employ three WDM models with thermal particle
masses of 0.8 keV, 1.3 keV and 2.6 keV, which are equivalent to
oscillation-produced Dodelson-Widrow sterile neutrino particle masses
of 3 keV, 6 keV, and 15.5 keV. Note that the 1.3 keV (thermal; 6 keV
sterile) case is equivalent in the structure formation cutoff scale of
the M2L25 model of \citet{Boyarsky:2008mt} \& \citet{Lovell:2011rd}.

Three separate issues require special attention when running these
simulations: 1) The dependence of the halo mass function at high
redshift on the chosen starting redshift, $\zini$; 2) Systematic
errors induced by the finite volume of the simulation; and 3)
Artificial haloes that emerge in WDM as a result of shot noise in
regimes where the underlying power spectrum is suppressed.  We discuss
each of these issues in turn.

Concerning the initial redshift $\zini$, recent advances in the
techniques used for numerical calculation of perturbations, such as
the second order Lagrangian Perturbation Theory \citep[2LPT; see, for
  example][]{Jenkins:2010}, have improved the convergence of
simulations using different $\zini$. Several groups
\citep[e.g.][]{Lukic:2007,Prunet:2008,Knebe:2009,Jenkins:2010,Reed:2013}
have worked to quantify the effect of $\zini$ on the final results of
cosmological simulations. These works stress the point that not using
2LPT algorithm leads to simulations that converge very slowly as the
start redshift is increased. In order to reduce as much as possible
any $\zini$ effect we have used the 2LPT algorithm incorporated in
MUSIC \citep{Hahn:2011} to generate all the initial conditions of our
simulations. Additionally, all the simulations presented in this work
use the same initial redshift ($\zini=125$). Therefore any systematic
effects associated with starting redshift will be present in all cases
and cancel when considering the ratios between the WDM and CDM halo
mass functions.

\begin{table}
\centering
\begin{tabular}{cccc}
\hline
$m_T$ & $m_{\nu_s}$ & $k_{1/2}$ [h/Mpc] & $k_{\rm max}$ [h/Mpc]\\
\hline
2.6 keV & 15.5 keV & 21.4 & 2.62\\
1.3 keV & 6.0 keV   & 9.47 & 1.28 \\
0.8 keV & 3.0 keV   & 5.24 & 0.764 \\
\hline
\end{tabular}
\caption{\label{tab:Transferfuncs} Warm dark matter models simulated
  in this paper.  The first column gives the thermal WDM mass.  This
  is the default model label we use throughout the work.  The second
  column gives the equivalent sterile neutrino mass.  The last two
  columns list the wave numbers where the transfer functions reach
  half the value of CDM and the wave numbers where the power spectrum
  is maximized, respectively.}
\end{table}

Systematic errors from the finite volume of the simulation box can be
divided into 3 categories: Shot noise, sampling variance, and lack of
power from modes larger than the simulation box. Shot noise is
especially important for the most massive haloes since only a few
exists in the simulation volume, and it generally decreases as
$1/\bar{n}V$ where $V$ is the simulation volume and $\bar{n}$ the
number count. However, for smaller halo masses, shot noise is dwarfed
by sample variance \citep{Hu:2002we}. The average density in the
simulation volume may happen to be an over- or under-dense part of the
universe, and since haloes are biased tracers of the density field,
this will lead to differences in the halo mass function. The best way
to correct this is to run independent samples of the underlying
density field (different seeds for the initial conditions), but this
comes at a considerable cost in terms of CPU hours. Alternatively, the
sampling variance can be estimated by analytic methods as given by
equation (4) in \citet{Hu:2002we}, with a Sheth-Tormen bias for
example. However, such a bias is based on fits to $\Lambda$CDM
simulations. It seems plausible that such a bias would not change
significantly if used in a WDM cosmology as it is primarily determined
by nature of halo collapse, but to avoid any complications with the
error estimate, we directly calculate the sample variance in the halo
mass function by the jackknife technique. We do not consider any
contributions to the halo mass function from scales larger than the
simulation box, since we are mostly interested in the low mass end,
and the simulated volume is significantly larger than the scale of
clusters at the redshifts of interest.

Finally, below a specific mass scale dependent on numerical
resolution, it has been well established that WDM produces artificial
haloes in simulations \citep{Wang:2007he,Angulo:2013sza}, an effect of
the shot noise due to the finite particle count. These haloes are
usually visible as regularly spaced clumps in the filaments of the
cosmic web, and they form below a mass scale proportional to $
m_p^{-1/3}$, where $m_p$ is the simulation particle mass. However,
force resolution also plays a role, and an excessive force resolution,
as compared to the mass resolution, can increase the number of
artificial haloes \citep{Angulo:2013sza}. \citet{Schneider:2013ria}
showed that the artificial haloes can be modeled by a power law
increase in the WDM halo mass function below the mass scale. Most
attention has been given to correcting the halo mass function for low
redshift, since contamination of the halo population is the largest
here, due to the fact that the artificial haloes have had more time to
form and accrete.

There have been no focused studies on artificial halo contamination is
at redshifts $z\gtrsim 5$.  In the results presented below, down to
the halo mass scale adopted for our completeness limit, we see little
if any indication of a low-mass upturn in our WDM halo mass functions;
such an upturn would be indicative of significant artificial halo
contamination.  Moreover, since any artificial haloes present would
provide an increase in the halo mass function (thus making WDM more
like CDM) ignoring them only makes our WDM constraints more
conservative.  In what follows, we have conservatively chosen to
ignore any corrections for artificial haloes in our catalogs.

Figure \ref{fig:snapshots} provides a qualitative depiction of the
differences inherent in WDM compared to CDM simulations.  Shown are
$10\times10\times6$ (comoving $h^{-1}$Mpc) slices of each of our
simulation volumes, centered on the most massive halo at a redshift of
$z=6$. On large scales the slices look similar, but on smaller scales
there is a clear lack of structure in the WDM models.

\begin{figure*}
\begin{centering}
\includegraphics{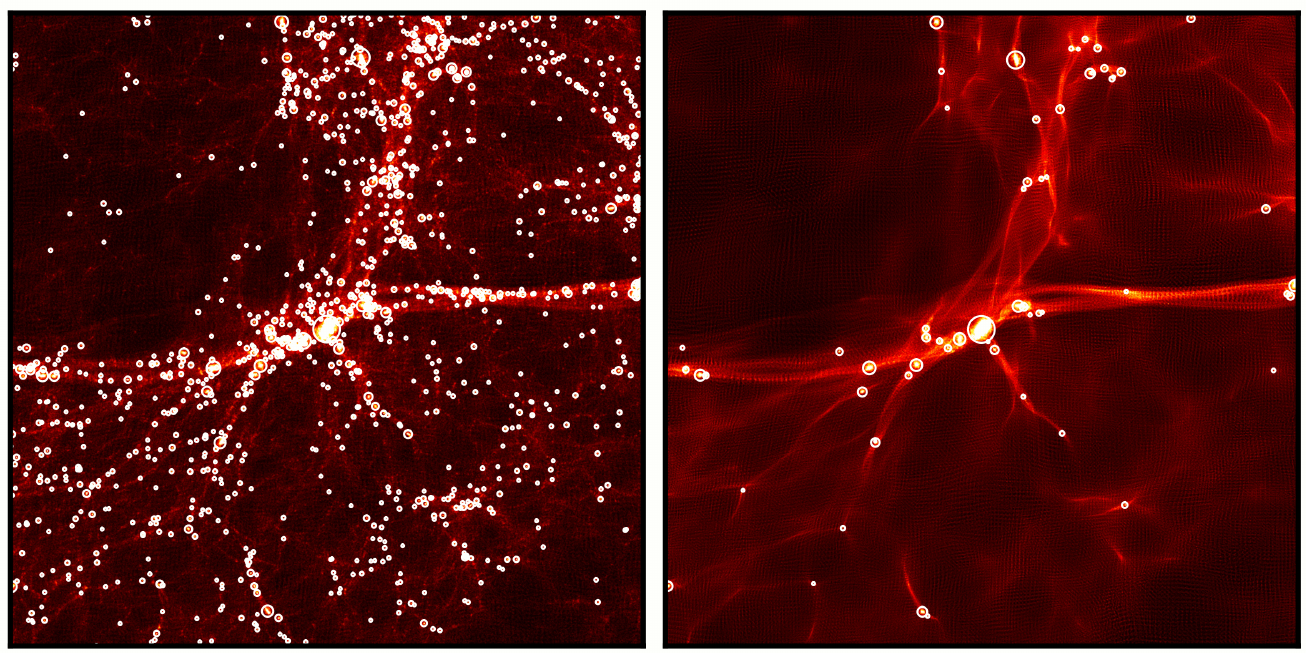}
\end{centering}
\caption{Simulation snapshots from CDM (left) and 0.8 keV WDM (right)
  overlaid with circles to indicate identified dark matter halos that
  are more massive than $3.4 \times 10^8 h^{-1} M_\odot$.  The size of
  the circle is proportional to the virial radius of each halo.  The
  CDM slice is filled with collapsed structure at z=6, while the WDM
  slice is largely devoid of collapsed halos that are massive enough
  for hydrogen cooling.  Note that artificial haloes would show up as
  regularly separated haloes in the filaments, suggesting that
  contamination by artificial haloes is likely negligible here.}
  \label{fig:snapshots2} 
\end{figure*}

\subsection{Halo catalogs}
We used the Amiga Halo Finder \citep[AHF,][]{Knollmann:2009} to
identify haloes in our simulations.  The halo mass $M_h$ used in this
work is calculated using the over-density ($\Delta_{\rm{vir}}$)
formula from \citet{Bryan:1998} for our cosmology at each specific
redshift. Note that our conclusions do not change when using different
over-density definitions, e.g. $\Delta_{200}=200 \rho_{\rm crit}$. As
explained above, to build our mass luminosity relation using the
abundance matching technique we took into account the merger history
of each halo and used its maximal mass obtained over its lifetime
$M_{\rm peak}$ instead of $M_h$. In any case, this correction turned
to be small due to the lack of substructure at high redshifts. We used
a requirement of at least 40 simulation particles to constitute a
halo, setting a halo mass completeness limit of $M_h = 3.4 \times 10^8
\, h^{-1}M_\odot$.

Compared to the density maps shown in Figure \ref{fig:snapshots}, the
differences between WDM and CDM become even more apparent when we
compare halo counts.  Figure \ref{fig:snapshots2} shows two of the
same density slices overlaid with white circles to indicate identified
dark matter halos more massive than our $M_h = 3.4 \times 10^8 \,
h^{-1}M_\odot$ completeness limit.  Circle sizes are proportional to
the virial radius of each identified halo.  The difference in
collapsed structures is striking between these two simulations.  For
example, the void in the upper left corner is completely empty of any
haloes in the 0.8 keV WDM run.

Figure \ref{fig:hmfs} provides a more quantitative demonstration of
the differences in halo abundances from model to model, where each
panel shows the cumulative dark halo mass function at redshifts $z=6$,
7, 8, and 13.  The CDM result (dotted line with shading) is in all
cases above the WDM models (solid lines with shading, as labeled).
\citet{Angulo:2013sza} found a suppression of the halo mass function
of the form\footnote{Strictly speaking \citet{Angulo:2013sza} has
  $\alpha=1$ fixed, however they also correct for artificial
  haloes. We find that keeping $\alpha$ as a free fitting parameter is
  necessary to provide reasonable fits, probably owing to a strong
  evolution with redshift. }
\begin{equation}
\frac{n_{\rm WDM}}{n_{\rm
    CDM}}(M)=\frac{1}{2}\left(1+\frac{M_1}{M}\right)^{-\alpha}\left[1+\text{erf}\left(\log\frac{M}{M_2}\right)\right].
\end{equation}
We have verified this expression provides a good fit to the WDM/CDM
abundance ratio for $z\lesssim10$, with decreasing accuracy with
increasing redshift.  In our simulations, at $10^9 M_\odot$, the 0.8
keV model is suppressed by more than an order of magnitude at all
redshifts relative to CDM.

As can be seen in the $z=13$ panel of Figure \ref{fig:hmfs}, no haloes
at all exist in 0.8 keV WDM model.  Indeed we find that no haloes have
formed before $z=12$ for 0.8 keV WDM and none before $z=15$ in the 1.3
keV model.  Detections at these epochs should be robust in the future
with JWST.  However, even current detections offer an interesting
test: the point with error bar ($2\sigma$) corresponds to the {\em
  lower limit} on the cumulative abundance of galaxies at those
redshifts, as set by the faintest galaxies observed in the HUDF
\citep[][]{Bouwens2007,McLure2012,Oesch2013}.  Its horizontal position
(corresponding halo mass) is based on the luminosity limit and our
adopted $M_h$-$L$ relation presented in the next section.
Importantly, the total abundance of galaxies at each redshift must be
above the data point shown (regardless of its horizontal positioning
on the plot).  One can see without any further analysis that the 0.8
keV WDM model will have trouble producing enough galaxies to match
{\em current} observations at $z > 8$; there are simply not enough
collapsed objects of any mass to account for the known galaxies at
this epoch. The viability of the other WDM models is not immediately
apparent from figure \ref{fig:hmfs} in itself, since the halo mass
function is not directly observed.
   
In order to provide a more precise connection with observations we
will need a mapping between halo mass and galaxy luminosity.  This is
a primary subject of the next section.

\section{Predicting Observables}
\label{sec:counts}

\subsection{Observed Luminosity Functions}
\label{sec:obslf}
We will normalize our predictions using observed high-$z$ galaxy
counts.  In doing so, we follow the literature and assume that
high-$z$ luminosity function is well characterized by a Schechter
function
\begin{equation}
\phi(L) \, dL =\phi_*\left(\frac{L}{L_*}\right)^\alpha\exp\left(-\frac{L}{L_*}\right) \, \frac{dL}{L_*}.
\end{equation}
Robust observations of luminosity
functions with measures of $\phi_*$, $L_*$, and $\alpha$ exist out to $z \sim
8$ \citep{Bouwens:2011xu,McLure2012,Schenker:2012vs} and 
current observations can provide constraints on the normalization (with other parameters fixed)
out to $z \sim 10$ \citep{Oesch:2013pt}.  

\begin{figure}
     \begin{centering}
           \includegraphics[width=\columnwidth]{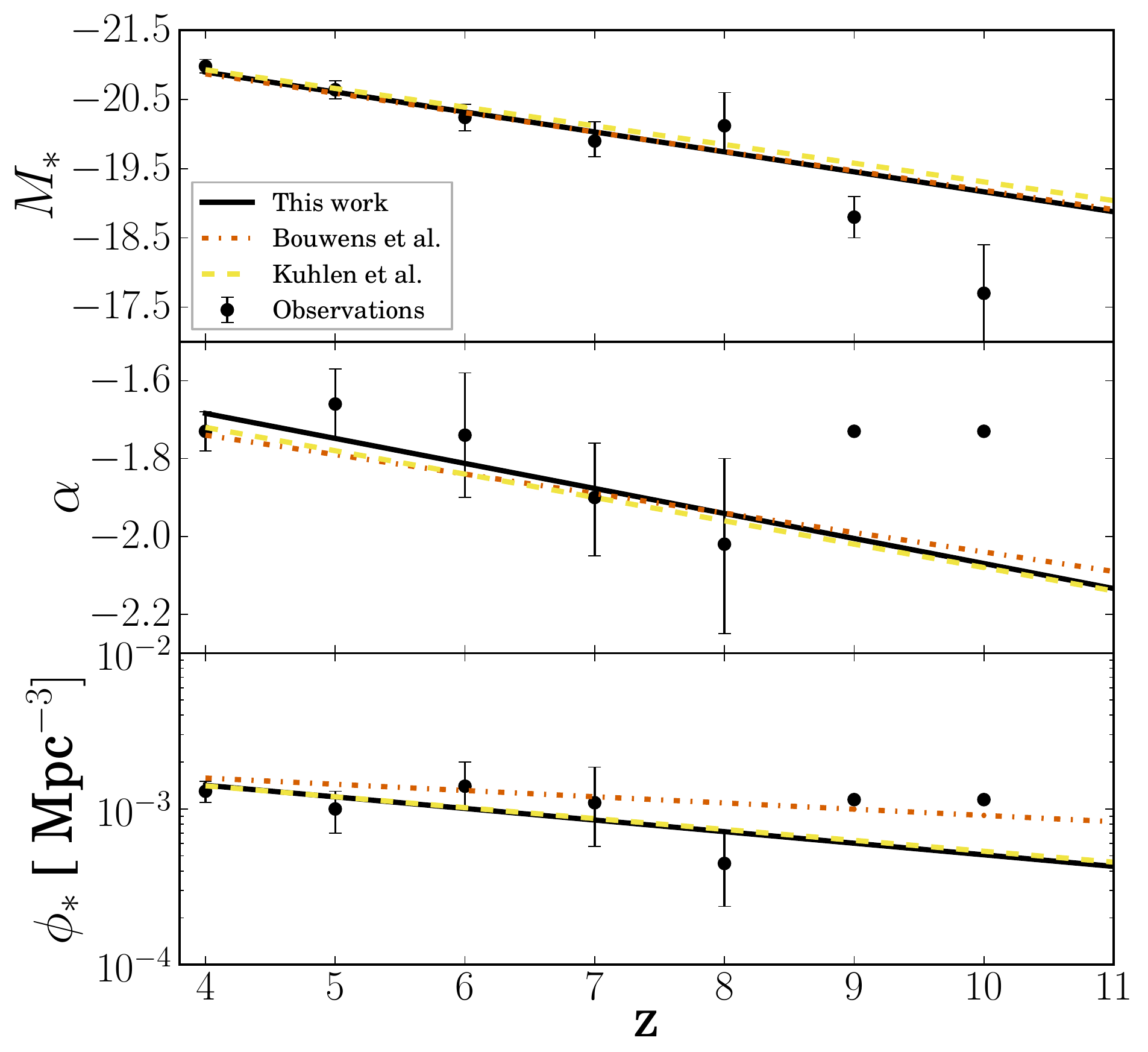}
     \end{centering}
     \caption{Shown are the fits for the Schechter function parameters
       $M_*,\alpha$ and $\phi_*$. The black symbols show observed
       values with quoted uncertainties
       \citep{Bouwens2007,McLure2012,Oesch2013}.  The lack of
       uncertainties for $\phi_*$ and $\alpha$ at $z=9$ and 10 is due
       to the fact that they were fixed in the estimation of
       $M_*$. Since there is a large correlation between $\alpha$ and
       $M_*$, these observations were not used for the fit. See
       \citet{Oesch2013} for further discussion. The solid black line
       is the fit used in this work, and we show other fits from
       \citet{Bouwens:2011xu} and \citet{Kuhlen:2012vy} for
       comparison.  Note that our fit uses newer observations from the
       2012 HUDF results than the comparison fits. For reference, our
       fits are $\log\phi(z)=-2.6-0.074z$, $M_{*}(z)=-22.1+0.29z$, and
       $\alpha(z)=-1.4-0.064z$.  }
   \label{fig:finkelfit}
\end{figure}

We parameterize the evolution of the luminosity function with redshift
by fitting quoted observational results for $\log\phi_*,L_*$ and
$\alpha$ and fitting them linearly as a function of $z$ from $z= 4-8$.
Figure \ref{fig:finkelfit} shows the fit used in this work in
comparison with fits from other authors. The data points used for this
fit (plotted) are taken from \citet{Bouwens2007} for $z=4-6$ and from
\citet{McLure2012} for $z=7 - 8$.  Points at higher $z$ (which assumed
fixed values for $\alpha$ and $\phi_*$) are shown for reference from
\citet{Oesch2013}.  Note that formally the luminosity density becomes
divergent if $\alpha<-2$, however, due to the introduction of a
minimum cutoff scale in halo masses in equation \eqref{eq:Mcool} this
is not a cause for concern. It is important to stress here that even
small changes in the fit parameters provides drastic changes in the
reionization history. This is especially true for changes to the faint
end slope $\alpha$ of the luminosity function
\citep{Bouwens:2011xu}. Future observations from the JWST can
hopefully much better constrain $\alpha$ at high redshift.

\begin{figure}
     \begin{centering}
           \includegraphics[width=\columnwidth]{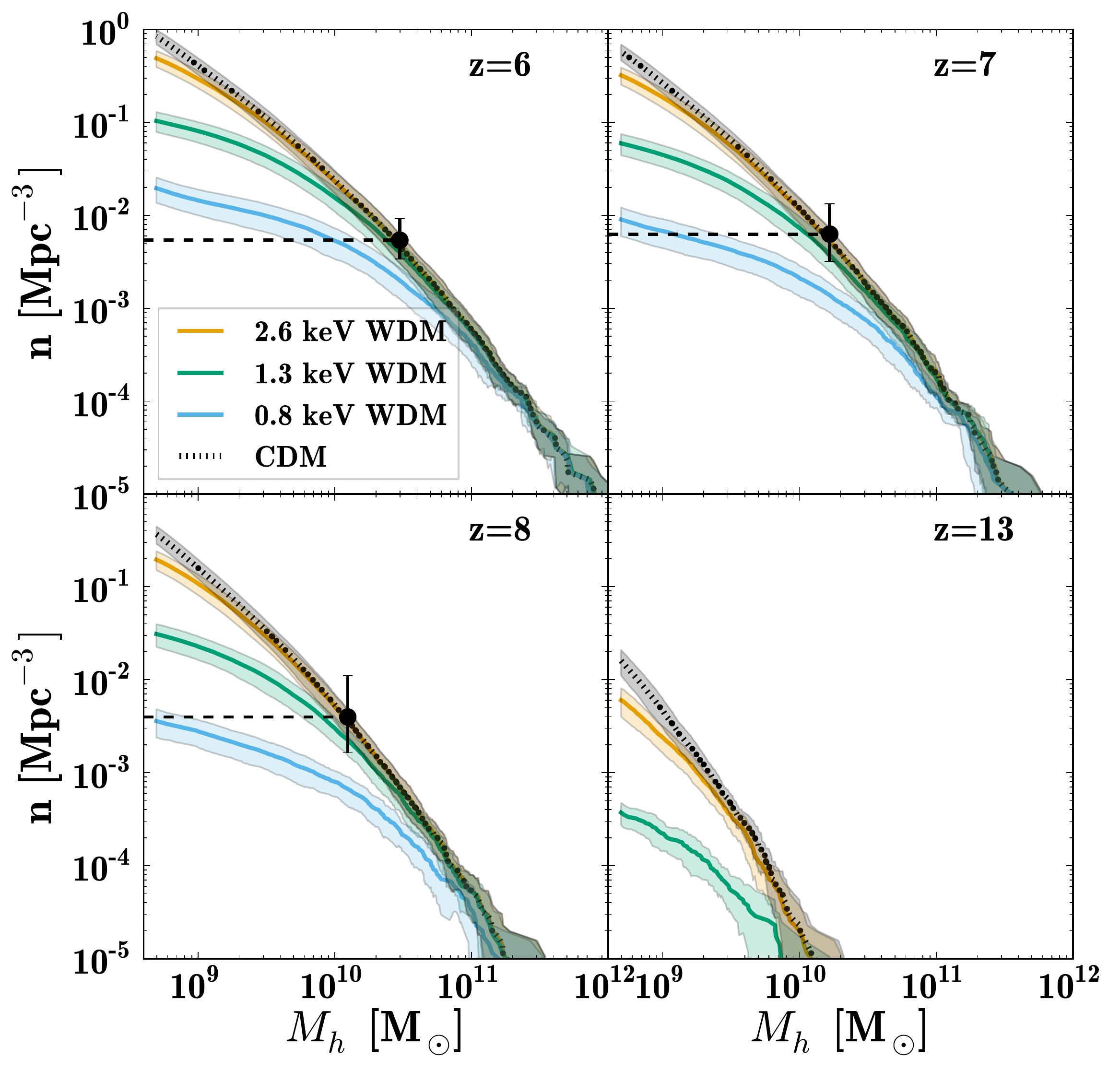}
     \end{centering}
     \caption{Shown are cumulative halo mass functions at selected
       redshifts for our CDM and WDM simulations. Notice how the
       suppression of WDM increases with $z$. At $z>12$ no dark matter
       haloes are identified at all in the 0.8 keV WDM model. The
       central lines denote the simulated halo mass function, while
       the shaded areas indicate jack-knife uncertainties. The
       horizontal dashed lines correspond to the known {\em lower
         limit} on the cumulative galaxy abundance at each redshift
       based on the faintest HUDF observations; assuming that galaxies
       reside in halos, any viable mode must produce a total abundance
       of halos above this line.  The point with 2$\sigma$ error bar
       is placed at the halo mass corresponding to the HUDF luminosity
       limit as inferred from abundance matching, discussed in Section
       3.  For constraints, we use the values of the galaxy luminosity
       functions observed and inferred from the halo abundance
       matching method, shown in figure \ref{fig:LFarr}.}
   \label{fig:hmfs}
\end{figure}

\subsection{Connecting Halos to Galaxies}
\label{sec:AM}
In assigning luminosities to dark matter halos we assume that brighter
galaxies reside in more massive halos and that the relationship
between halo mass and galaxy luminosity is monotonic, following the
same relation for all dark matter models.

Fundamentally we rely on the abundance-matching technique
\citep{Kravtsov2004,Vale:2004yt}, which defines the relationship
between halo mass and galaxy luminosity (or alternatively stellar
mass) by equating the cumulative number density of halos to the
cumulative number density of galaxies observed.  The power of this
approach is that the observed luminosity function is fully reproduced
(at least down to luminosities where the observations are complete or
to where the matching is performed) while sweeping all uncertainties
galaxy formation physics under the rug.  In principle other halo
parameters could be used as the rank order of choice \citep[e.g.,
  maximal circular velocity][]{TrujilloGomez:2010yh}.  Recently,
Behroozi and collaborators have argued that halo mass is the most
robust variable to use for these purposes \citep{Behroozi2013c}.

\begin{figure}
     \begin{centering}
           \includegraphics[width=\columnwidth]{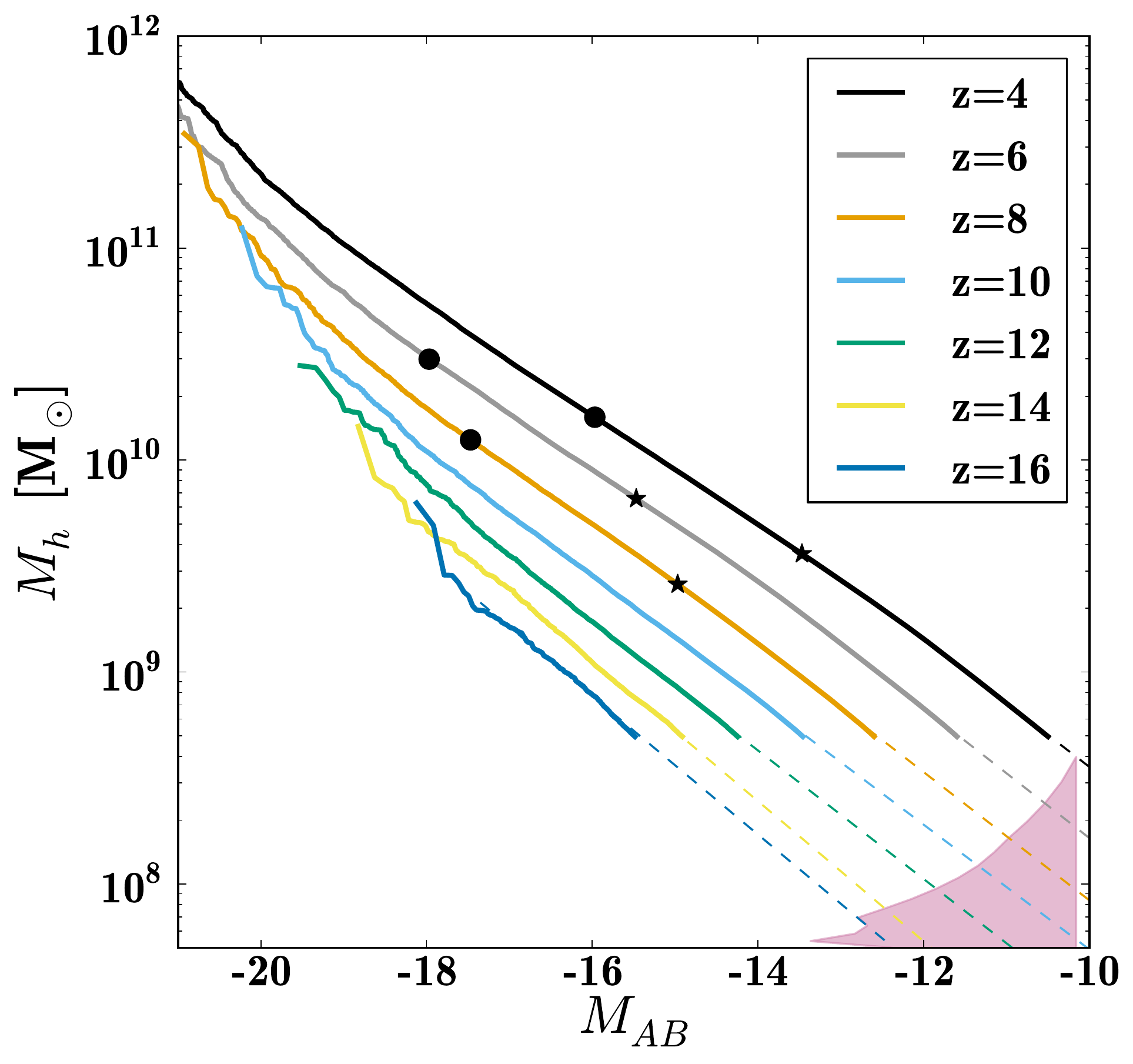}
     \end{centering}
     \caption{Shown is the relationship between halo mass and galaxy
       magnitude adopted in this work.  The solid lines are derived
       from our simulations using abundance matching as described in
       the text.  They are truncated at the point where are dark
       matter halo catalogs become incomplete. The dashed lines are
       power law extrapolations, fit to the solid lines at the faint
       end as $\log M_h = C + b \, M_{AB}$, with resulting slopes of
       $b = -0.35$ to $-0.30$.  The shaded area is an estimate of the
       cooling limit for halos, below which galaxies cannot form
       efficiently, via equation \eqref{eq:Mcool}. The circles
       indicate the current HUDF magnitude limit, the asterisks are
       the expected JWST limits.}
   \label{fig:AM}
\end{figure}

Specifically, we set the relationship between halo mass $M_h$ and UV
luminosity $L$ at different redshifts z via
\begin{align}
n_{\rm CDM}(M_h,z) &= \Phi_{\rm gal}(L,z) \, , \label{eq:CDMabundance} 
\end{align}
where $n_{\rm CDM}$ is the cumulative dark halo count in CDM and
$\Phi_{\rm gal}$ is the cumulative luminosity function as given by the
Schechter function fits discussed in section \ref{sec:obslf}.  The
resultant relationships at various redshifts are plotted in Figure
\ref{fig:AM}.  As can be seen, we are fundamentally assuming that the
relationship between halo mass and galaxy luminosity obeys a power-law
at faint magnitudes (normalized at the bright end by observations)
with $M_h \propto L_{AB}^{a}$, $a \approx 0.75$ (or $\log M_h \sim
-0.3 M_{AB}$).

Our fundamental assumption is that the halo mass-luminosity mapping is
a power law at faint magnitudes. Furthermore we are conservatively
assuming that the relationship between halo mass and galaxy luminosity
is the same in CDM and WDM (obeying near power-law behavior at faint
luminosities).  This approach demands that all models match the
observations at the bright end (where halo counts overlap), and makes
the assumption that there is no special break (towards more efficient
galaxy formation) in the luminosity-halo mass relation in WDM for
small halos. This is conservative because there is no reason to expect
that WDM halos will be more efficient at making galaxies than CDM
halos.  Indeed, WDM halos collapse later and have had less time to
form stars, so we might expect them to be less efficient at forming
stars than their CDM counterparts.

\citet{Herpich:2013} performed hydrodynamical simulations of several
different WDM cosmologies. They found only slight differences between
the stellar masses of the different dark matter models they
considered, and the difference they did see was towards less efficient
formation in WDM as discussed above. For the most extreme case,
comparing a 1 keV WDM model to CDM, they found a ratio of
$M_{\star,\rm CDM}/M_{\star,\rm WDM}\sim 2$ in stellar masses at
$z=0$. Therefore, the main differences in the star formation histories
are produced at late times, and therefore this relatively small effect
is more reduced at the high redshifts of interest for this work. Small
variations in the stellar feedback implementation have a much greater
impact on the final stellar mass of a galaxy than WDM particle
mass. While the results of \citet{Herpich:2013} are based on low-$z$
simulations it seems reasonable to expect the same to be true at
higher redshift. Appendix \ref{sec:appA} further elaborates on using
the CDM halo catalogue as the fiducial model.

The assumption that the power law behaviour for the faint end of the
halo mass-luminosity mapping translates into a suppression of the
luminosity function in the corresponding WDM model. This suppression
is then exactly equal to the ratio of the halo catalogues at the given
mass scale, which is an aggressively conservative assumption of a
cancelation of two disparate physical mechanisms. Furthermore, this
requires that the WDM luminosity functions diverge from a pure
Schechter fit. We stress that several other processes could also cause
the luminosity function to diverge from a power law, however, since no
such divergence is yet observed, this translates into constraints on
the nature of dark matter.

Before moving on we note that while the general abundance-matching
approach has proven successful and robust at reproducing galaxy
properties in the low redshift universe, it is less well tested at
higher redshifts.  For example, the scatter at fixed halo mass appears
to be fairly minimal at low-$z$ \citep{Behroozi2013c} and the
relationship between halo mass and luminosity is well described by a
power law for faint systems \citep{Moster2013,GK2013}.  At high
redshift however, the relationship between halo mass and UV luminosity
could in principle exhibit significant scatter, though a power-law
relationship for the smallest galaxies appears consistent with the
data \citep{Behroozi2013c}.  We adopt a strict one-to-one relationship
between halo mass and UV luminosity as a starting point in
investigating the expected differences between CDM and WDM on galaxy
counts in the high-z universe.  Because we are looking at differential
effects between the two models, driven by the declining number of
low-mass halos in WDM, we anticipate that this approach provides a
fair starting point, though it would be useful to extend this approach
to more complicated mappings in the future.\footnote{Notably, we take
  the inherent scatter in the halo mass-luminosity mapping into
  consideration.}

\subsection{Reionization}
With predictions for luminosity functions in hand, we can directly
connect those to expectations on reionization.  Star forming galaxies
at $z\gtrsim 6$ are the primary candidate for the main process driving
the reionization of the intergalactic medium (IGM). Changes in the
abundance of early galaxies therefore translates into different
reionization histories.

The reionization process is a tug-of-war between ionizing radiation
from short-lived massive stars and atomic recombination in the IGM. In
terms of the volume filling fraction of ionized hydrogen $Q_\HII$ this
is captured in the differential equation \citep{Kuhlen:2012vy}
\begin{align}
\frac{dQ_\HII}{dt}&=\frac{\dot{n}_{\rm ion}}{\langle
  n_H\rangle}-\frac{Q_\HII}{\bar{t}_{\rm rec}}\ ,  \label{eq:reion_de}
\intertext{where $\dot{n}_{\rm ion}$ is the creation rate of ionizing
  photons, and $\langle n_H\rangle$ is the comoving density of baryons}
\langle n_H\rangle&=X_p\Omega_b\rho_{\rm crit} \ ,
\intertext{and
  $\bar{t}_{\rm rec}$ the mean time of HII recombination}
\bar{t}_{\rm rec}&=\frac{1}{C_\HII \alpha_B(T_0)\langle n_H\rangle
  (1+Y/4X)(1+z^3)} \ .
\end{align}
Here $\alpha_B$ is the case $B$ recombination
coefficient\footnote{Commonly used case $A$ and $B$ definitions
  differentiate mediums that allow the Lyman photons to escape or that
  are opaque to these lines (except Lyman-$\alpha$) respectively. Case
  B is most appropriate for this reionization calculation.}, $T_0$ is
the IGM temperature and $X$ and $Y=1-X$ are the primordial hydrogen
and helium abundances respectively. Since recombination is not
isothermal and uniformly distributed, the gas clumping factor
$C_\HII=\langle n_H^2\rangle/\langle n_H\rangle^2$ is also introduced
to quantify the effects these approximations have. Allowing a fraction
of $f_{\rm esc}$ of the produced ionizing photons to escape the gas
clouds where the massive stars are born, the injection of UV photons
into the IGM is given by the differential luminosity function $\phi$
down to a limiting luminosity $L_{\rm lim}$
\begin{equation}
\left.\dot{n}_{\rm ion}=\zeta_{\rm ion}f_{\rm esc}\int_{L_{\rm
    lim}}^\infty L\phi(L)dL=f_{\rm esc}\zeta_{\rm
  ion}\rho_{UV}\right|_{L_{\rm lim}} . \label{eq:nion}
\end{equation}
Here $\zeta_{\rm ion}$ is a parameter converting the galactic UV
luminosity to ionizing photon luminosity, or more precisely the
amount of Lyman continuum photon emission per 1500 \AA\ unit UV
luminosity density. Note that $f_{\rm esc}$ and $\zeta_{\rm ion}$ are
completely degenerate parameters, as reionization is only sensitive to
the product. Any change in one of these parameters could be attributed
to the same relative change in the other.

Critical in this analysis is what value to assign to the limiting
luminosity in equation \eqref{eq:nion}, that is, the minimal UV
luminosity expected possible from early galaxies. Naively, one might
expect no such lower limit. However, to capture the hot primordial gas
needed for star formation, a sufficiently deep potential well is
required. This effectively puts a lower bound on the possible UV
luminosities (and therefore halo masses) due to
photo-evaporation. Furthermore, star formation can only take place
once the hot gas has cooled sufficiently, and this introduces a
limiting mass threshold below which stars cannot form. In this work we
only consider the cooling limit for halo masses, and we use the
parameterization adopted by \citet{Sobacchi:2013ww}
\begin{equation}
M_{\rm cool} = 10^8\left(\frac{1+z}{10}\right)^{-3/2} M_\odot.
\label{eq:Mcool}
\end{equation}
The shaded red area in Fig.~\ref{fig:AM} shows an estimate of the
region where galaxy formation is suppressed owing to this cooling
limit, effectively mapping the halo mass limit to a luminosity cutoff.
Based on this, we will explore cutoff magnitudes between $M_{AB}=-10$
and $M_{AB}=-13$ in what follows.

An important constraint on reionization comes from the Thomson optical
depth to the CMB,
\begin{equation}
\tau_e=\int_0^{z_{R}} \frac{c(1+z)^2}{H(z)}Q_\HII \sigma_T \langle
n_H\rangle\left(1+\eta(z) Y/4X\right).
\end{equation}
Here $z_{R}$ is the redshift of recombination, $\sigma_T$ is the
Thomson cross section and $\eta=1$ when Helium is singly ionized and
$\eta=2$ when Helium is doubly ionized after $z\lesssim 4$.

In this work, we do not include an evolution in the reionization
parameters. \citet{Kuhlen:2012vy} found that for their best fit
scenario, evolution in the limiting luminosity alone is not enough to
match both Ly$\alpha$ constraints and reionization constraints, and
the data provides no conclusive evidence for an evolution in any
case. Evolution in $f_{\rm esc}\zeta_{\rm ion}$ (resulting perhaps
from evolution in the stellar initial mass function) may be more
plausible.  We will present our results with different values of
$f_{\rm esc}\zeta_{\rm ion}$ and limiting luminosity, but they will
remain fixed with redshift. Evolution in the clumping factor $C_\HII$
may be expected, but no definitive determination of its evolution
exists. For example, \citet{Finlator:2012gr} presents a detailed
analysis of the evolution, suggesting that the clumping factor rises
from $C_\HII<1$ for $z>10$ to $C_\HII \sim 3.3$ at $z\sim 6$. It
should be noted that our model equation \eqref{eq:reion_de} does not
include the detailed distribution of hydrogen where some dense clumps
reionize later than less dense clumps. This simplification is likely
inaccurate in the final phase of reionization around $z\sim 6$,
however we expect it to be an appropriate approximation on average for
higher-$z$ and in the large cosmological volumes of interest here.

In what follows we adopt the reionization parameters $C_\HII=3$,
$\zeta_{\rm ion}=10^{25.3} \text{ergs}^{-1} \text{Hz}$, $X_p=0.75$,
$T_0=2\times 10^4 \text{K}$ and $\alpha_B=1.6\times
10^{-13}\text{cm}^3/\text{s}$. The escape fraction and limiting
luminosities vary, and will be indicated in the relevant figures.

\section{Results}
\label{sec:Results}
Here, we explore the constraints on WDM models by direct number
counting and the inferred reionization history, and illustrate how
future galaxy count surveys can improve these constraints on WDM
models.

\begin{figure}
     \begin{centering}
           \includegraphics[width=\columnwidth]{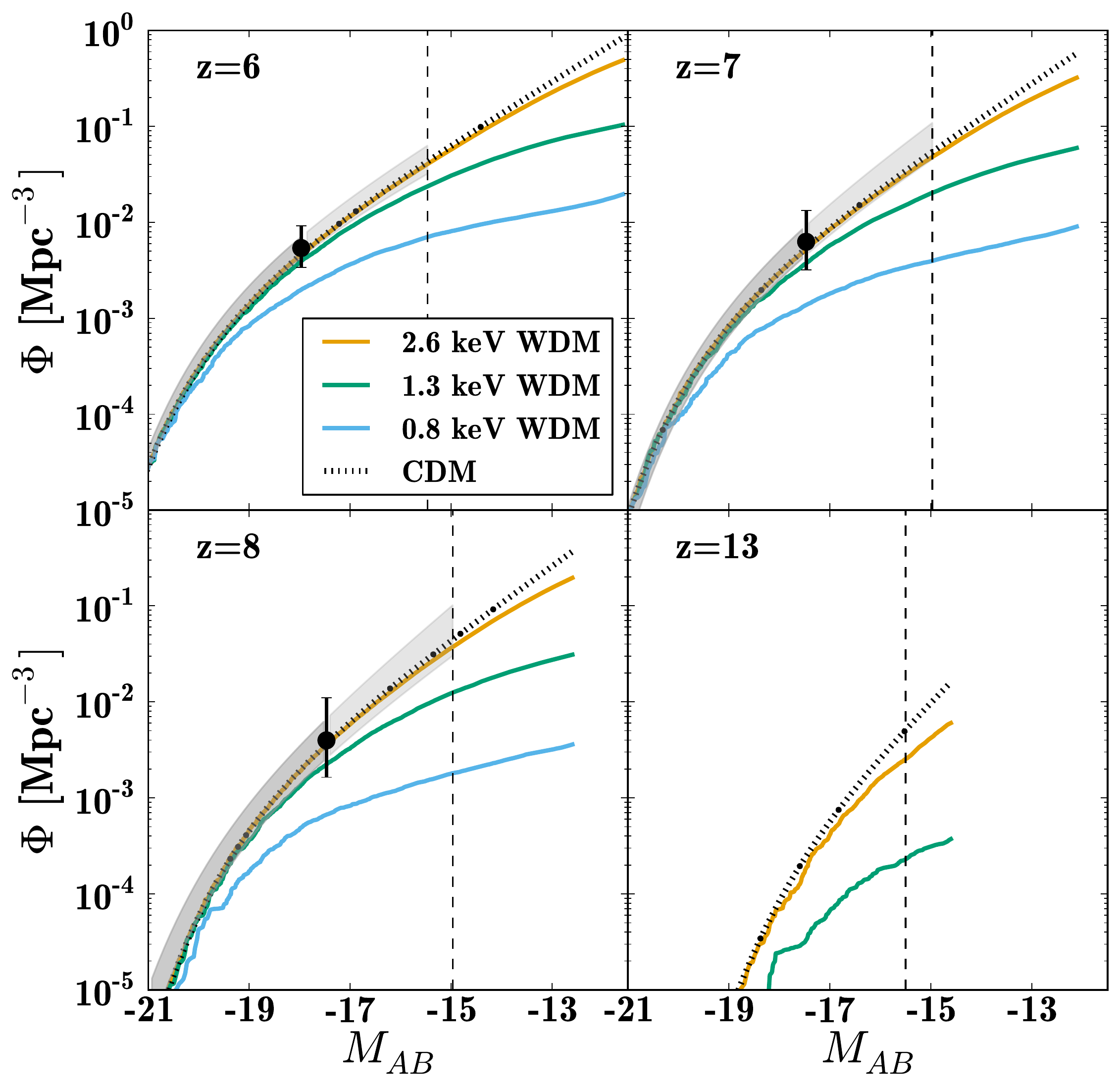}
     \end{centering}
     \caption{Shown are cumulative luminosity functions for our CDM
       and WDM models at various redshifts (comoving volumes). In each
       panel, the symbol with error bar marks the observed cumulative
       count at the limit of published HUDF luminosity functions with
       2$\sigma$ uncertainties shown.  The shaded bands bands
       correspond to 1$\sigma$ uncertainties, and the vertical line
       marks the approximate reach of a hypothetical deep field
       observation with JWST. For redshifts where the luminosity
       function has been observed by HUDF the JWST limit has been
       assumed to be 2.5 magnitudes fainter. For $z=13$ a limit of
       -15.5 has been assumed (see \citet{Windhorst:2005as}).  The 0.8
       keV WDM model is heavily disfavored by current observations,
       and the 1.3 keV model is marginal.  At redshift $z=13$, JWST
       observations will likely be able to rule out 1.3 keV WDM and
       perhaps be sensitive to 2.6 keV.}
   \label{fig:LFarr}
\end{figure} 

\subsection{WDM Constraints from Galaxy Counts}
\label{sec:LFres}

As explained in section \ref{sec:AM}, given our (conservative)
assumption of the star formation efficiency being the same in the
different dark matter models, deviations in the implied WDM luminosity
function from the observed Schechter fit translates into constraints
on the WDM model.

Figure \ref{fig:LFarr} shows the implied luminosity functions for CDM
(dotted black) and each of our WDM models (solid, colors indicated).
The symbol with error bar is the known (observed) cumulative count of
galaxies at the faint HUDF limit, with errors indicative of the
$2\sigma$ uncertainty calculated as in Fig.~\ref{fig:hmfs}. The shaded
band corresponds to a $1\sigma$ uncertainty of the observed best fit
Schechter function (seen in figure \ref{fig:finkelfit}), extrapolated
down to an approximate JWST deep field limit (indicated by the
vertical line). We see clearly here that the 0.8 keV WDM model (solid
cyan) is strongly disfavored by current observations of the galaxy
luminosity functions. The 1.3 keV model, while currently consistent
with observations, demonstrates significant deviations from CDM at
$z=13$ at magnitudes observable with JWST.  Deep galaxy counts at this
and earlier epochs may be even sensitive to 2.6 keV WDM.

We quantify how much the different models are disfavored with a
$\chi^2$ test,
\begin{equation}
\chi^2=\sum_i\left(\frac{\Phi-\Phi_{\rm obs}}{\sigma}\right)^2 \ .
\end{equation}
Here $\Phi_{\rm obs}$ is the abundance at the faint-end limit from
observed luminosity functions \citep{Bouwens2007,McLure2012,Oesch2013}
and $\sigma$ is the error on the simulated luminosity function, which
is given by the jackknife error on the halo mass function at the
corresponding abundance.\footnote{We use the jackknife errors instead
  of the errors on the observed luminosity function, since they are
  larger (shown in Fig.~\ref{fig:LFarr}). The $\chi^2$ test here is
  approximate, but our conclusions are robust regarding the models'
  consistency. } The CDM luminosity function is a fit to the redshift
evolution of the Schechter function parameters (shown in
Fig.~\ref{fig:finkelfit}) based on current observations, and not the
actual quoted fits at each redshift, and this produces a small but
non-zero $\chi^2=2$ for the CDM case with 5 degrees of freedom
(corresponding to the observations from $z=4$ to $z=8$) from the
luminosity functions (85\% consistency).  The $\chi^2$ for the 2.6,
1.3 and 0.8 keV models are 2.27, 14.4 and 372, respectively, with
probabilities for these models at getting the observed luminosity
functions of 81\%, 1.3\% and $\ll 10^{-10}$. Therefore, the 1.3 keV
WDM model is disfavored at approximately 98.6\% C.L. (2.2$\sigma$),
and the 0.8 keV WDM model is disfavored at very high significance,
$>10\sigma$.  Note that the statistical methods in the modeling and
constraints here do not reflect systematic uncertainties in the halo
abundance matching method. Because we arrive at a consistent and
smooth power law relation for the abundance matching, the systematic
effects are likely small, though difficult to quantify.

Had we considered WDM artificial halos and the fact that the star
formation efficiency is slightly lower in WDM than in CDM the
constraints would improve slightly, since there would then be fewer
and less luminous galaxies. However, these effects are not likely to
be very important at redshifts $z\geq 4$, and are \textit{a priori}
unknown.

\begin{figure}
     \begin{centering}
           \includegraphics[width=\columnwidth]{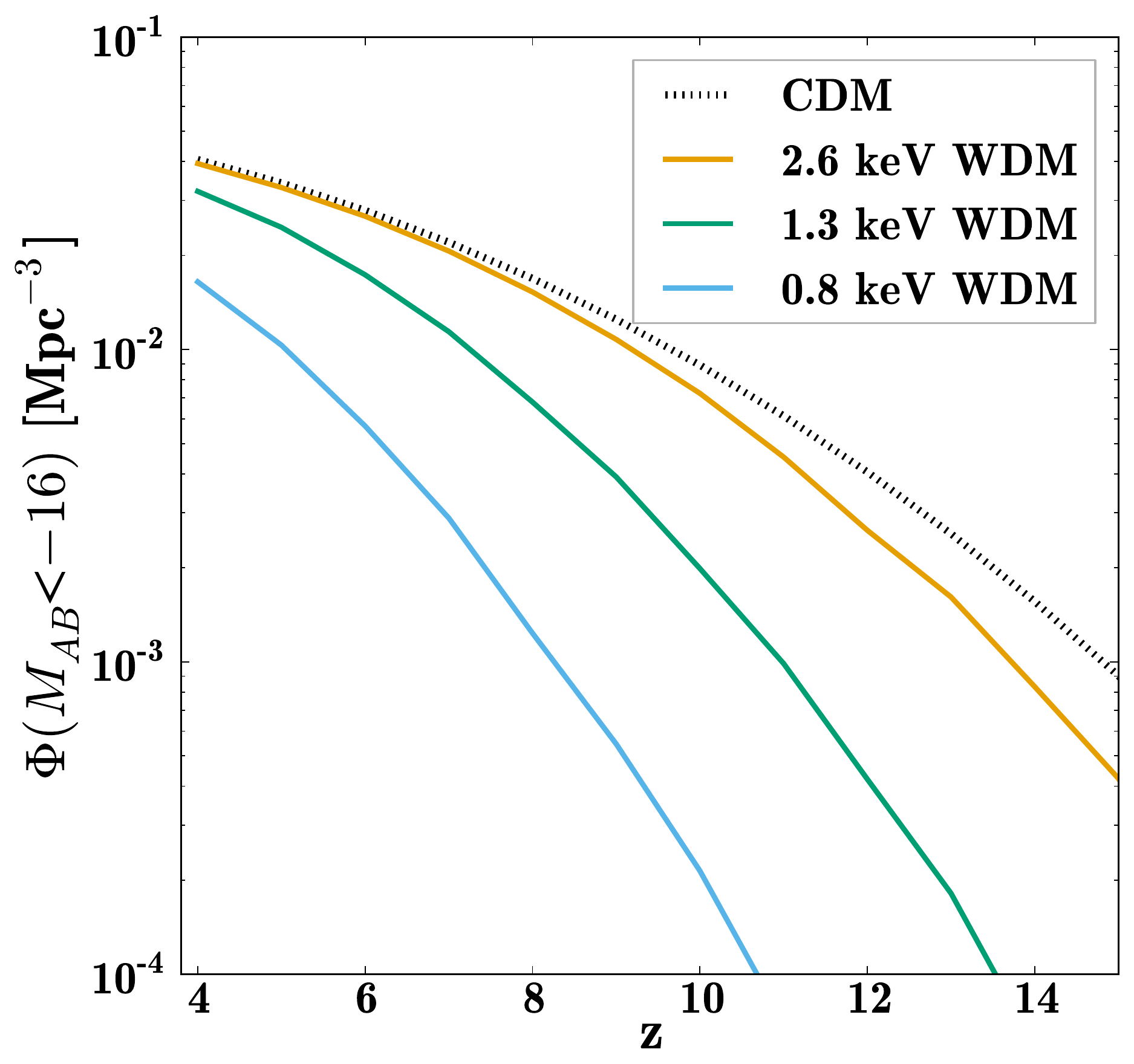}
     \end{centering}
     \caption{Predicted number density of galaxies brighter than
       $M_{AB}=-16$ as a function of redshift for our CDM and WDM
       models.  JWST should be capable of detecting galaxies of this
       brightness across the redshift range plotted, and perhaps be
       sensitive to differences between 2.6 keV WDM and CDM at $z >
       12$.  }
   \label{fig:Mcutplot}
\end{figure}

Faint galaxy counts at even higher redshift will be particularly
sensitive to WDM models.  We demonstrate this in Figure
\ref{fig:Mcutplot}.  Here we show the cumulative number density of
galaxies brighter than $M_{AB} = - 16$ as a function of redshift for
each of our models.  The differences between CDM and WDM are
significant, especially for the lower mass WDM cases.  Deep JWST
observations should be sensitive to galaxy detections at least this
faint out to $z=15$, and therefore will provide a direct probe of the
small-scale power spectrum by counting galaxies.

\begin{figure}
     \begin{centering}
           \includegraphics[width=\columnwidth]{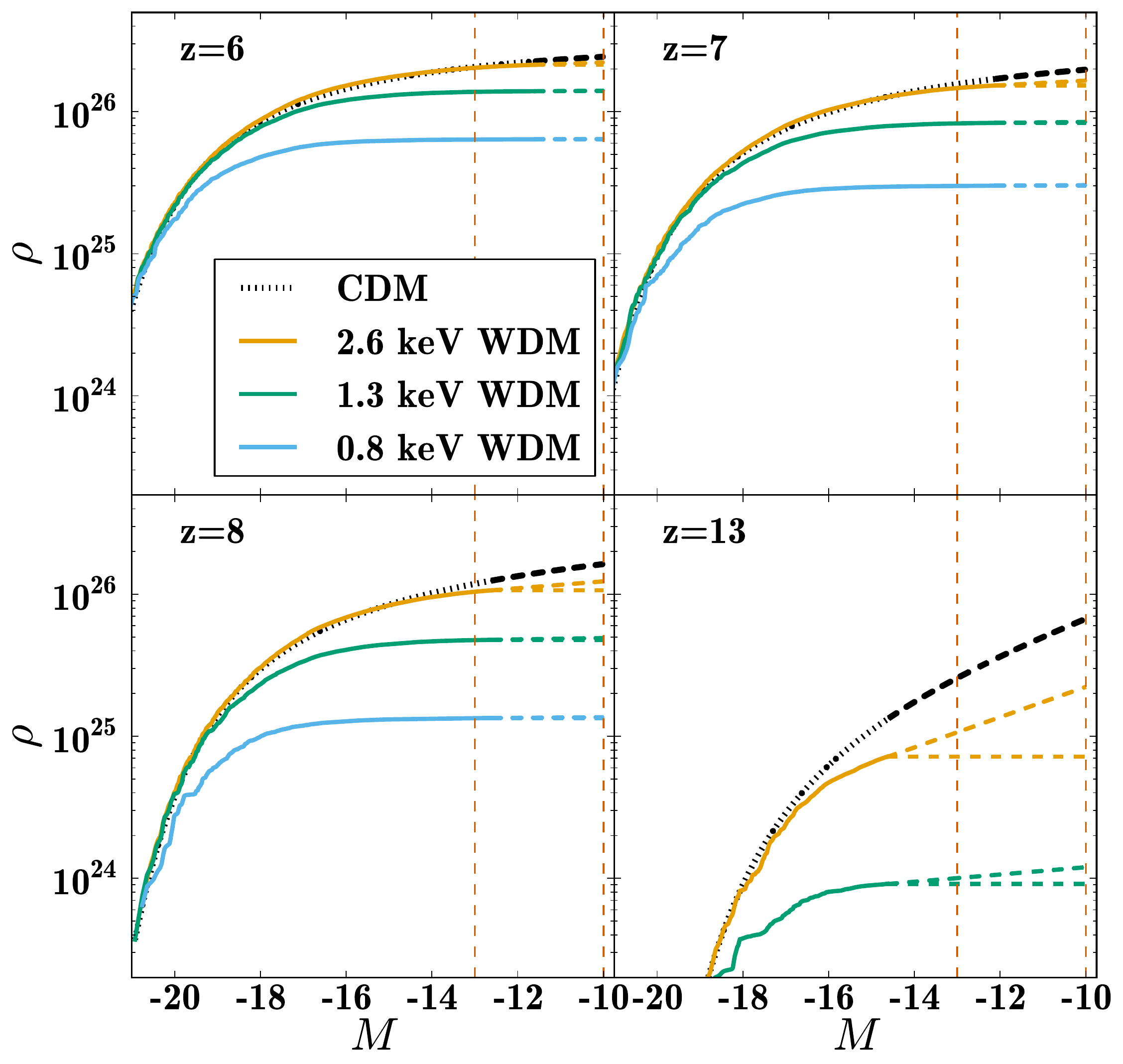}
     \end{centering}
     \caption{The cumulative UV luminosity density (in units erg/s
       Hz$^{-1}$ Mpc$^{-3}$) as a function of magnitude cutoff at
       selected redshifts for the dark matter models considered.  The
       vertical lines mark the two cutoff scales we consider in this
       paper as plausible for extending the galaxy luminosity
       function.  The change from solid to dashed lines occurs at the
       magnitude corresponding to the resolution limit of the
       simulation, beyond which we rely on extrapolations (dashed) to
       predict faint galaxy contributions.  For a given WDM model, the
       upper dashed lines extend the best-fit power-law of the
       resolved function.  The lower dashed line marks the constant
       value at the faintest simulated point, as would be expected if
       the WDM halo mass function drops dramatically beyond this
       point. These two extremes bracket reasonable expectations.  }
   \label{fig:Lumdens}
\end{figure}  

\subsection{WDM Constraints from Reionization}
\label{sec:Lumres}
Fundamental to the ability of galaxies to reionize the universe is the
production rate of ionizing photons, $\dot{n}_{\rm ion}$, which is
proportional to the total UV luminosity density, $\rho_{UV}$, coming
from these sources (see equation \ref{eq:nion}).  For an underlying
galaxy luminosity function with a steep faint-end slope $\alpha$, the
total luminosity density implied will be sensitive to the assumed
faint-end cutoff used to calculate $\rho_{UV}$.  Figure
\ref{fig:Lumdens} shows the luminosity density in our models as a
function of faint-end cutoff at selected redshifts. Because WDM models
have flatter faint-end luminosity function slopes, the total
$\rho_{UV}$ is less sensitive to the faint-end cutoff, i.e. the
implied cumulative $\rho_{UV}$ values flatten relative to CDM at
fainter magnitudes.  Importantly for our considerations, WDM
predictions for reionization will be less sensitive to the adopted
faint-end cutoff than CDM, owing to the lack of small galaxies in
these models.

The points where the lines change to dashed in Fig.~\ref{fig:Lumdens}
mark the resolution limit in the simulations. The two WDM dashed lines
bracket the following extreme cases: one, a power law fit to the faint
end, and, two, the constant value at the faintest point resolved in
the simulations. The actual luminosity density would be somewhere
between these two extremes, though for the 0.8 keV and 1.3 keV models
the difference is negligible. All our analysis utilizes the power law
extrapolations to get conservative estimates.

\begin{figure}
     \begin{centering}
           \includegraphics[width=\columnwidth]{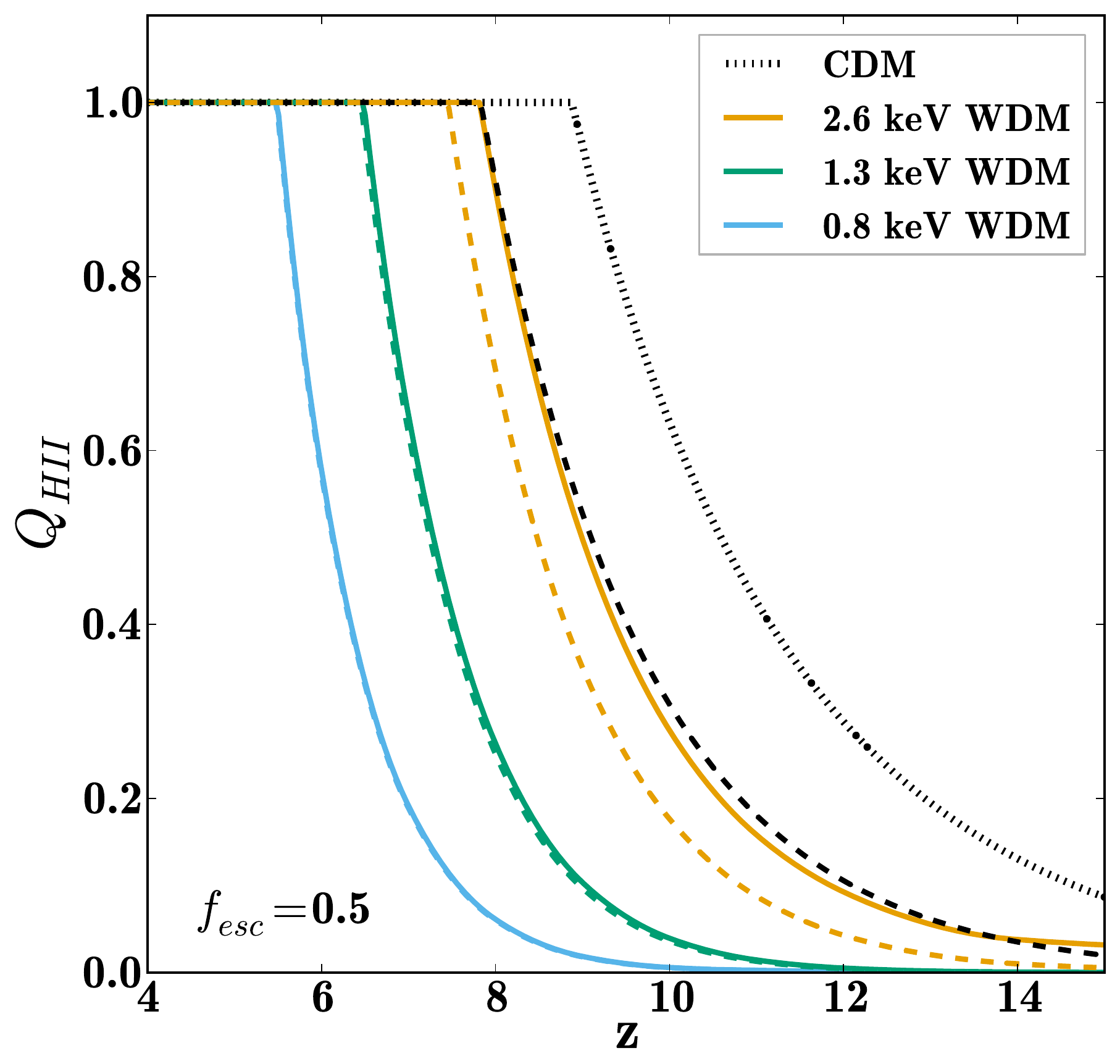}
     \end{centering}
     \caption{Buildup in volume filling fraction of ionized hydrogen
       as a function of redshift for our CDM and WDM models assuming
       $f_{\rm esc}=0.5$ and with limiting integration magnitudes of
       $M_{AB}=-10$ (fiducial lines, as in caption) and $M_{AB}=-13$
       (dashed).  The WDM models produce more rapid buildups at later
       times compared to CDM.  The 0.8 keV WDM model does not complete
       reionization by $z\sim 6$ (as seems to be required by
       observations) regardless of the assumed faint-end cutoff. }
   \label{fig:Qplot}
\end{figure}

With the luminosity density in hand, the reionization history can now
be determined by virtue of equation \eqref{eq:reion_de}.
Fig.~\ref{fig:Qplot} presents the volume filling fraction $Q_\HII$ as
a function of redshift for two choices of limiting magnitude in
calculating the luminosity density.  The fiducial line types (shown in
the legend) correspond to a limiting magnitude of $M_{AB}=-10$ while
dashed lines cut off at a brighter limit of $M_{AB}=-13$. We used the
initial condition $Q_\HII=0$ at $z=20$ and integrated forward in
time. We choose an optimistic escape fraction of $f_{\rm esc} = 0.5$,
higher than assumed in both \citet{Robertson:2013bq} and
\citet{Kuhlen:2012vy}, and therefore more conservative with respect to
WDM model constraints since we use the same $\zeta_{\rm ion}$ as in
\citet{Kuhlen:2012vy}. \citet{Robertson:2013bq} uses a lower
$\zeta_{\rm ion}$, but instead assumes an almost constant luminosity
density at the cutoff scale at high redshift.

All of the models in Figure \ref{fig:Qplot} except 0.8 keV WDM have
completed reionization by $z \sim 5.8$ as required by results inferred
from the kinematic Sunyaev-Zeldovich effect and CMB polarization
observations \citep{Zahn:2011vp}. For the 0.8 keV model shown,
reionization is complete at $z = 5.5$.  In general, the WDM cases
produce a more rapid late-time buildup of ionized hydrogen due to the
high redshift suppression of haloes. It can also be seen that the
difference between CDM and 2.6 keV WDM is larger when the fainter
limiting magnitude is used, simply because the difference between the
models is much larger here.  Of course, these results are sensitive to
the escape fraction.  For example, if an escape fraction of $f_{\rm
  esc}=0.2$ is used for the 1.3 keV model full reionization is not
reached until $z = 5.4$ (not shown on figure), so fairly high escape
fractions seem to be required for 1.3 keV to reach full reionization
by $z\sim6$.

\begin{figure}    
 \begin{centering}
\includegraphics[width=\columnwidth]{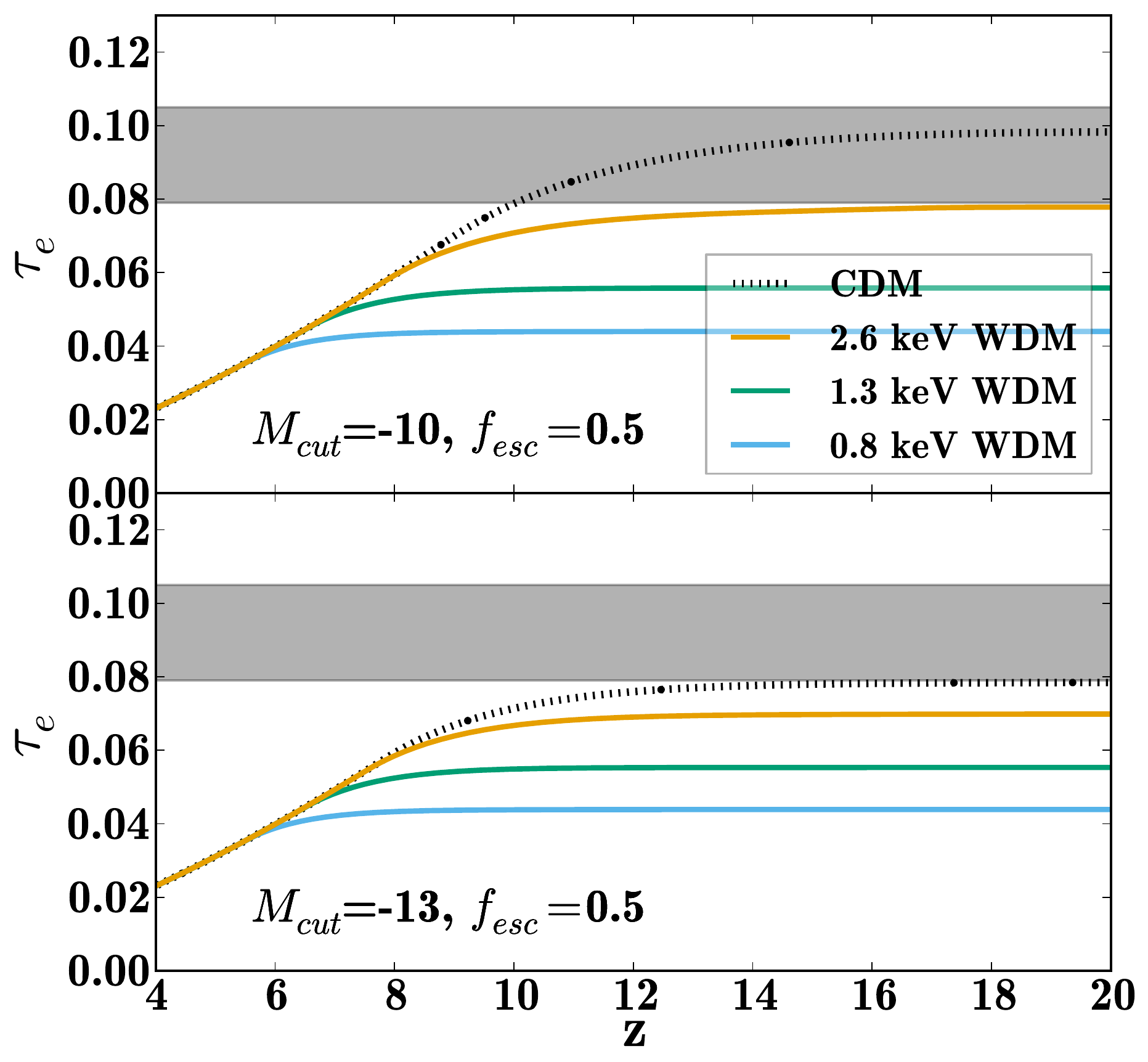} 
\end{centering}
\caption{The electron-scattering optical depth of the CMB predicted
  for our CDM and WDM models, with the contribution shown cumulatively
  as a function of redshift.  The top panel shows results for an
  assumed luminosity function cutoff at $M_{AB} = -13$ and the bottom
  panel extends this cut to very faint luminosities $M_{AB} = -10$.
  In all case we assume $f_{\rm esc} = 0.5$ and our fiducial value of
  $\zeta_{\rm ion}$.  The bands are the 68\% confidence limit on the
  most recent {\it Planck} results \citet{Ade:2013zuv}. Note that none
  of the WDM models reach within the 68\% confidence band from {\it
    Planck}.  }
     \label{fig:tau1}
\end{figure}

Another important probe of reionization is the integrated optical
depth of electron scattering from the CMB.  The shaded bands in Figure
\ref{fig:tau1} show the CMB optical depth range of $0.092\pm0.013$
from the most recent {\it Planck} results \citep{Ade:2013zuv}.  The
lines show predictions for the optical depth as contributed as a
function of redshift for our WDM and CDM models assuming an escape
fraction $f_{\rm esc}=0.5$, with the two panels corresponding to
different limiting magnitudes.  Interestingly, with this choice of
(fairly high) escape fraction, none of our WDM models can reproduce
the measured optical depth, and even CDM requires a luminosity
function extrapolation to a very faint limiting magnitude.  This is
consistent with the findings of \citet{Robertson:2013bq}.

\begin{figure}    
 \begin{centering}
\includegraphics[width=\columnwidth]{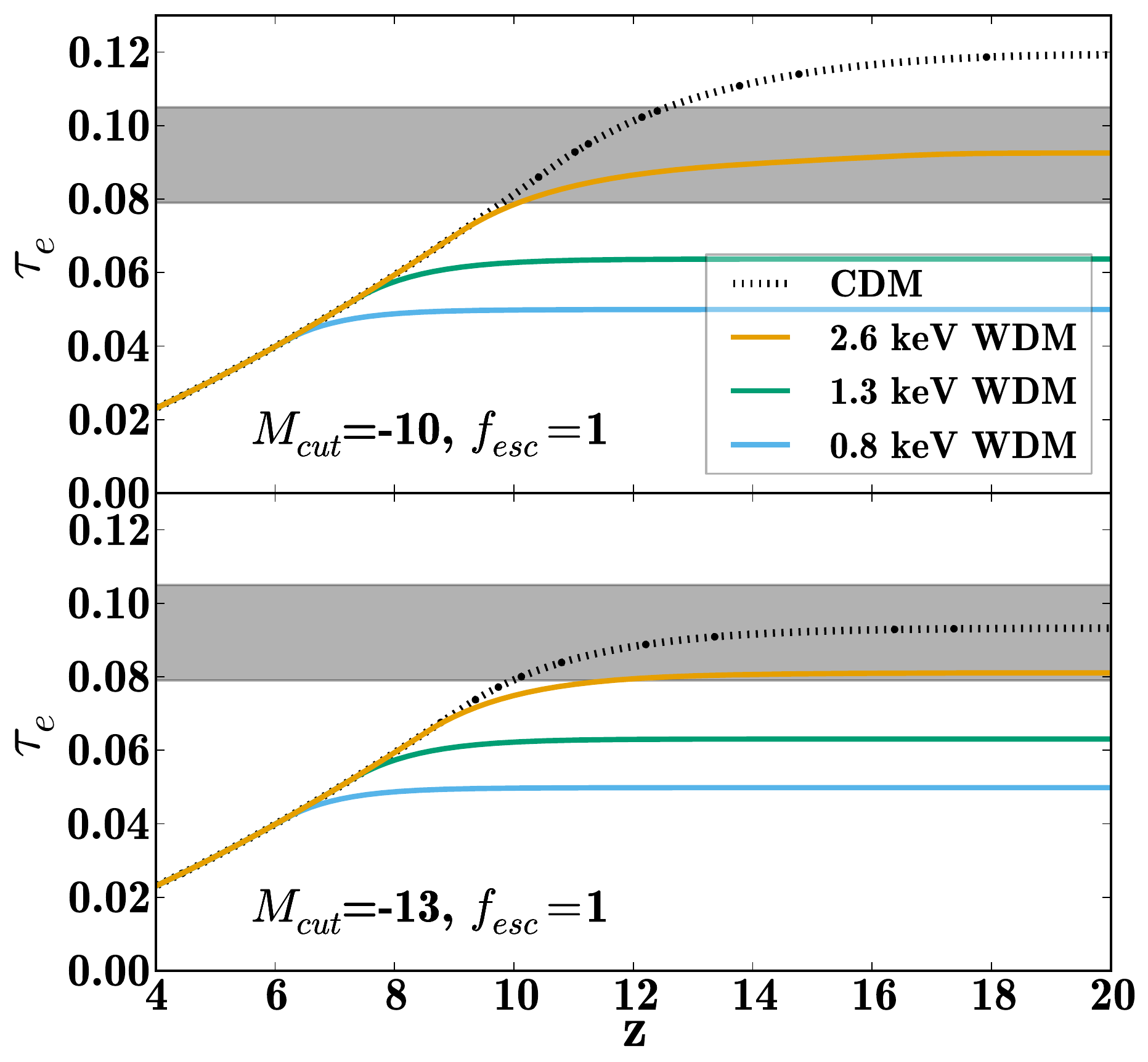}
\end{centering}
\caption{Electron-scattering optical depth as in Fig.~\ref{fig:tau1}
  except now assuming $f_{\rm esc} = 1.0$. Even with fairly extreme
  assumptions, neither of the two lightest WDM models are able to
  reach the 68\% confidence range (bands) reported by {\it Planck}.}
     \label{fig:tau2}
\end{figure}

In Fig.~\ref{fig:tau2}, we show results for the optical depth, now
assuming $f_{\rm esc}=1$.  In this case, the 2.6 keV model can
reproduce the Planck value, though a fairly faint limiting magnitude
seems to be required, even in this extreme case.  Unsurprisingly, CDM
severely overshoots the optical depth with these (rather high)
reionization parameters. It is noteworthy that neither of the low mass
WDM models can reproduce the Planck optical depth within its 68\%
confidence interval, even with very optimistic choices.  If these WDM
models are to be viable in the face of reionization constraints, they
would require either significant contribution to the ionizing flux
from non-stellar sources, or a significantly larger combination of
$f_{\rm esc}\zeta_{\rm ion}$ than what is currently believed to be
realistic. A smaller $C_\HII$ could also help in this respect.  Future
observations will likely better constrain these parameters.

\section{Discussion}
\label{sec:Discussion}
Future measurements of the luminosity function of faint galaxies at
high-$z$, particularly those from JWST, and possibly with HST via the
Frontier Fields initiative, will significantly improve the sensitivity
to WDM models and the halo mass cutoff effects presented here. For a
direct number count comparison the mass resolution of our simulations
is sufficient to connect with observations down to plausible detection
limits with JWST.

In contrast, our results on reionization specifically for the 2.6 keV
model would improve with greater mass resolution. This is because of
our conservatively approximated increasing faint end luminosity
density function, which likely flattens at low luminosities in this
model, but remains unresolved in our simulations.  This improvement
can also be made, though to a lesser degree, for the lower WDM
particle mass models. Our results for these particle mass models
indicate that the luminosity density reaches an approximately constant
level at the faint end, and thus the ionizing flux will remain
constant at fainter luminosities. Deep observations with JWST
certainly will much better determine the faint end slope $\alpha$ of
the luminosity function. The reionization history is highly sensitive
to this parameter. A future analysis could possibly circumvent the
uncertainty stemming from fitting the evolution of the Schechter
parameters, instead relying on direct observations of $\alpha$ for
most redshifts.

We have neglected any evolution in the escape fraction with
redshift. Na\"ively, one might expect the escape fraction to decrease
with redshift since the overall density scales as $(1+z)^3$. However,
observations seem to indicate the opposite: the escape fraction
increases with redshift. \citet{Mitra:2012av} found that $f_{\rm
  esc}\sim 0.06$ at $z\sim 6$, and increases to at least $f_{\rm
  esc}\gtrsim 0.146$ by $z\sim 10$. This could be caused by an initial
mass function for star formation favoring high mass stars in the early
universe. Alternatively, \citet{Ferrara:2012zy} proposes a mechanism
where mini haloes close to the cooling limit contributes appreciably
to the ionizing flux, but their contribution diminishes over time due
to feedback mechanisms. Since these small haloes have a relatively
larger escape fraction the overall escape fraction decreases with
time. This mechanism is especially interesting from a WDM perspective,
since the lower abundance of small haloes directly counteracts
this. The evolution of the escape fraction remains uncertain, and a
typical constant value in previous work was $f_{\rm esc}\sim 20\% $,
and our conservative choice of $f_{\rm esc} > 0.5$ will very unlikely
overestimate the ionizing flux.

Non-stellar processes can potentially contribute to the reionization
history. Quasars might play an important role at high redshift,
although current results seem to indicate the contribution from
quasars is sub-dominant \citep{Volonteri:2009ck}. For example,
\citet{Willott:2009wv} found that at $z\sim6$ the ionizing flux from
quasars is 20-100 times lower than the what is needed for continued
reionization. X-ray emission by black holes may also contribute
appreciably.  \citet{Ricotti:2003vd} and \citet{Ricotti:2004xf}
analysed a pre-ionization contribution from a top-heavy initial mass
function for the population III stars. These population III stars then
collapse into black holes and subsequently accrete at nearly the
Eddington limit. Accretion onto the black holes could partially
reionize the IGM, although the primary effect would be heating the
IGM. The population III phase is rapidly self-limiting due to
pollution by heavy elements and pair instability supernovae causing
strong outflows, and thus the metal-poor population III stars needs to
collapse into black holes. This pre-ionization phase is then followed
by a period of stellar reionization.

\citet{Bovill:2010bz} and \citet{Katz:2012nd} present a mechanism
where the bulk of the star formation in the first dwarf galaxies
happened in proto-globular clusters that were subsequently tidally
stripped from the dwarf galaxy. Since the tidal stripping of globular
clusters in the halo outskirts precede the stripping of the dark
matter halo this can break the assumption of the abundance matching
technique since the dwarf galaxies would be stripped of much of their
luminous matter. Consequently, the mass-luminosity mapping would no
longer be strictly monotonic, and the scatter on any mass-luminosity
relation would likely increase. Importantly, for mass scales larger
than the masses of these dark dwarf galaxies, however, the abundance
matching would still give a meaningful average mass-luminosity
mapping. In any case, the abundance matching we employ here would be
an upper bound on the luminosity of a dwarf galaxy and, therefore, our
approach is highly conservative.

In the case of sterile neutrino WDM, it has been shown that the
effects of the radiative decay of sterile neutrino WDM to X-ray
photons may catalyze the formation of $\rm H_2$ and star formation
\citep{Biermann:2006bu}. This does not affect the results presented
here, because in this case the sterile neutrino WDM cosmology is
constrained to produce the same observed high-$z$ luminosity functions
from which the reionization history is inferred. The only method by
which such radiative decays would enhance the reionization rate is if
they preferentially enhanced star formation for small mass halos below
the luminosity function cutoff magnitude. We are aware of no mechanism
in the literature that would produce such an enhancement for low mass
halos. Moreover, this X-ray photon catalizatoin process would enhance
star formation preferentially in metal free low mass halos that rely
on $H_2$ to cool. More massive halos which are involved in
reionizatoin are insensitive to this mechanism as they can cool by
Ly$\alpha$ line emission and are more likely to be metal
rich. Therefore, since halo formation is suppressed at these low
masses, we believe our results apply for the case of sterile neutrino
WDM, with the 3 keV and 6 keV mass scales disfavored at $>10\sigma$
and $98.6\%$ C.L., respectively.

Recall that the 1.3 keV (thermal; 6 keV sterile) model we have
considered corresponds to model discussed by \citet{Lovell:2011rd} as
a solution to the too-big-to-fail problem \citep{BoylanKolchin:2011dk}
and the M2L25 model studied in \citet{Boyarsky:2008mt}.  We have
demonstrated that this model is disfavored at $98.6\%$ C.L. by direct
galaxy counts at high redshift and is unable to reproduce the CMB
optical depth even with extreme assumptions about the escape fraction.

\section{Conclusion}
\label{sec:Conclusion}
We have shown that the Lyman-break technique for galaxy surveys at
high redshift can provide a direct method for constraining the nature
of dark matter and its clustering at small scales, with sensitivity to
the structure formation suppression present in WDM models.  We have
analyzed CDM and WDM cosmological simulations in order to test WDM
models using the luminosity function observations at high-$z$ as well
as a new analysis of cosmological reionization limits. Given the
assumptions that the luminosity function of a $\Lambda$CDM universe is
modeled by a Schechter function down to faint magnitudes and that the
mass-luminosity relation of galaxies is independent of the dark matter
model employed, we have modeled the luminosity function for several
dark matter models to analyze the sensitivities to WDM dark matter
models.

Using an approximate $\chi^2$ test of the faint end of the luminosity
function, direct number counts of galaxies significantly disfavors a
0.8 keV WDM model at greater than 10$\sigma$, and a 1.3 keV model is
disfavored at approximately 98.6\% C.L. (2.2$\sigma$). Further, with
highly optimistic values for the parameters that translate high
redshift galaxy luminosity to ionizing flux, the 0.8 keV and 1.3 keV
model are inconsistent with the CMB optical depth at greater than
$68\%$ C.L. Furthermore, for the conservative case of a limiting
luminosity of $M_{AB}=-13$, a 2.6 keV WDM model is only marginally
consistent with the 68\% confidence region of the optical depth from
{\it Planck}.  Wherever possible, we have used conservative values on
parameters, making WDM behave more like cold dark matter.  For this
reason we feel confident concluding that neither the 0.8 keV or 1.3
keV models are consistent at more than 68\% C.L. with reionization,
even with the large uncertainty on the reionization process.

We expect upcoming deep surveys with JWST (and possibly HST via the
Frontier Fields) to be able to reach luminosities and redshifts that
can fully discern between a CDM model and a 1.3 keV model by direct
number counts.  Even 2.6 keV WDM might prove discernible if the
observations are deep enough.  Additionally, if the constraints on
reionization parameters are improved, a 2.6 keV WDM model can be
distinguished from cold dark matter by its different reionization
history. The study of galaxy formation and reionization in the
high-$z$ universe adds a complementary and competitive probe to the
nature of dark matter.

\section*{Acknowledgments}

We thank useful discussions with John Beacom and Richard Ellis. We
would also like to thank the referee Massimo Ricotti for helpful
feedback on the paper. KNA is partially supported by NSF CAREER Grant
No. PHY-11-59224.  JO and JSB were supported by grants from the NSF
and NASA.

\appendix
\section{Abundance matching with warm dark matter}
\label{sec:appA}

\citet{Herpich:2013} found that the star formation in low-$z$
Milky-Way-like galaxies is slightly suppressed in WDM
cosmologies. This seems to fit well with what one would expect: the
small-scale cut-off in the WDM transfer function postpones the
formation of dwarf galaxy halos, and therefore the potential wells
that act as seeds for the first galaxies are shallower. The effect is
relatively small, only a factor of 2 for their most extreme WDM model
at $z=0$.

\begin{figure}
     \begin{centering}
           \includegraphics[width=\columnwidth]{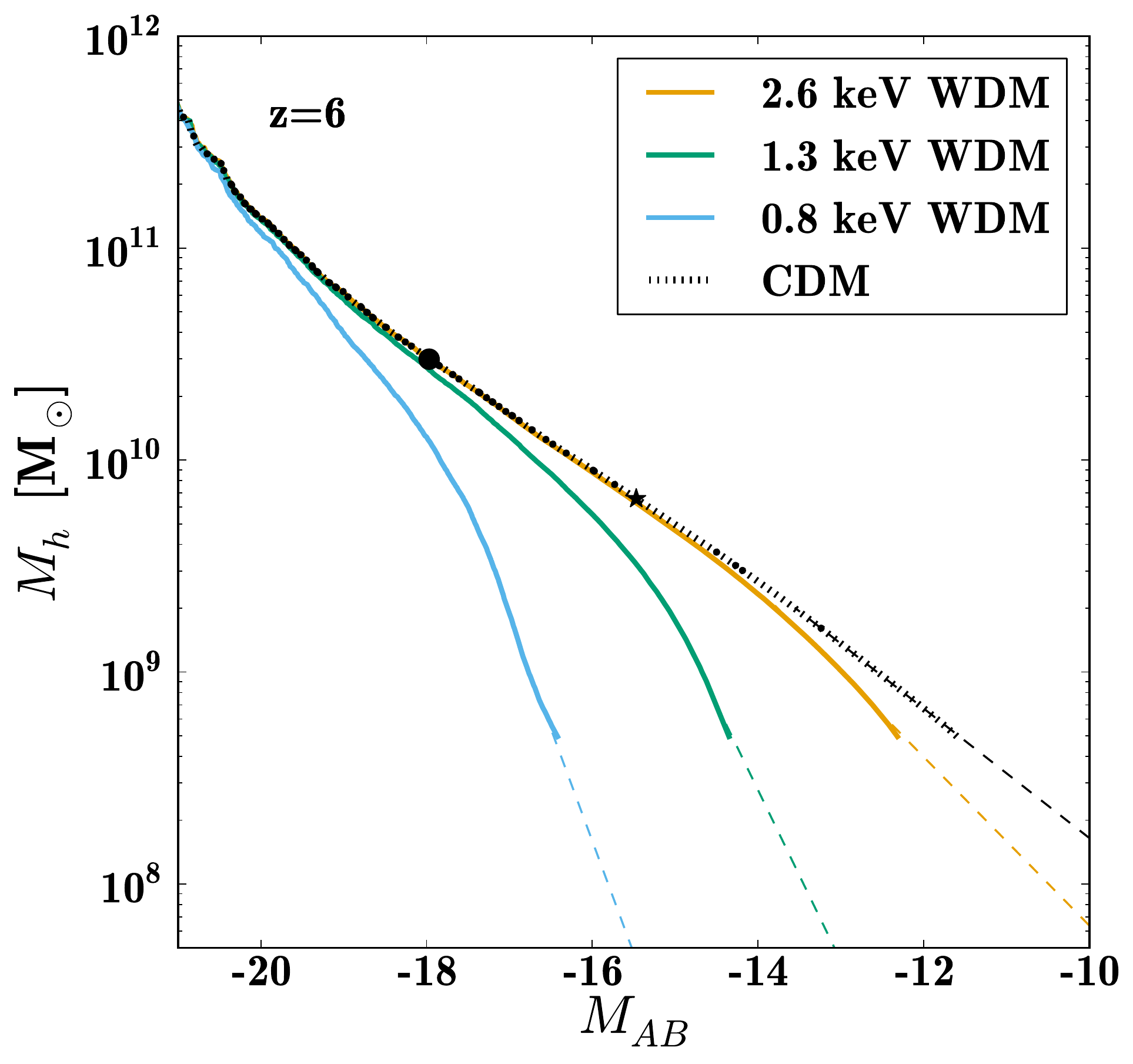}
     \end{centering}
     \caption{Abundance matching utilising the different halo
       catalogues. Abundance matching with the CDM catalogue gives a
       power law down to faint magnitudes. The dashed lines are power
       law extrapolations to the faint end. Clearly, the faint end in
       the WDM models diverge from the CDM power law towards more
       efficient star formation: lower mass halos have a larger
       luminosity relative to CDM. However, the WDM models must have
       roughly the same or a sligtly lower star formation rate than
       CDM, hence our assumption of the CDM power law behaviour is a
       very conservative estimate. The circle indicate the current
       HUDF magnitude limit, the asterisk is the expected JWST
       limits.}
   \label{fig:AM_WDM}
\end{figure}

The star formation efficiency for different dark matter models can
readily be inferred from abundance matching. Figure \ref{fig:AM_WDM}
shows the halo mass-luminosity relation by using the different dark
matter halo catalogues. For a fixed mass, a WDM halo is seen to be
more luminous than a CDM halo, and thus WDM would, unsurprisingly,
need to have an enhanced star formation efficiency relative to CDM in
order to match observations. This is counter-intuitive, and more
importantly contradicts the low-$z$ results of \citet{Herpich:2013}. A
realistic WDM halo mass-luminosity mapping would give a slightly lower
star formation efficiency: that is, a flatter slope than CDM in figure
\ref{fig:AM_WDM} instead of a steeper slope. In our analysis we
therefore conservatively assume that the halo mass-luminosity is a
power law. Therefore, a halo mass can uniquely be mapped to the same
luminosity independent of WDM model.

\bibliographystyle{mnras}
  \bibliography{ref}

\end{document}